\documentclass[review, times]{elsarticle}

\usepackage[inline]{trackchanges} 
\usepackage{url} 
\usepackage{lineno}
\usepackage{amsmath}
\usepackage{commath}
\usepackage{amssymb}
\usepackage{amsbsy}
\usepackage[utf8]{inputenc}
\usepackage{nomencl}
\usepackage{bbold}

 \newcommand{\bu}{\mathbf{u}}
 
\journal{Advances in Water Resources}
\graphicspath{ {./images/}}
 \usepackage{graphicx,caption}
 \usepackage[export]{adjustbox}
 \usepackage{wrapfig}

\begin{document}

\title{Pore network models to determine flow statistics and structural controls in variably saturated porous media}

\author[inst1,inst2]{Ilan Ben-Noah}
\author[inst1]{Juan J. Hidalgo}
\author[inst1]{Marco Dentz}

\affiliation[inst1]{Institute of Environmental Assessment and Water Research (IDAEA), Spanish National Research Council (CSIC), Barcelona, Spain.}
\affiliation[inst2]{Department of Environmental Physics and Irrigation, Institute of Soil, Water and Environmental Sciences, The Volcani Institute, Agricultural Research Organization, Rishon LeZion, Israel}

\begin{frontmatter}

\begin{abstract}
We study the abilities of pore network models of different complexity to determine the flow statistics and their structural controls in partially saturated porous media. The medium permeability and hydraulic tortuosity are the basic parameters for upscaling flow problems from the pore to the Darcy scale. They represent average flow properties. 
However, upscaling and predicting dispersion and anomalous solute transport from the pore to the continuum scale requires knowledge of the velocity distribution, not only its mean values. 
Considering four different network models of increasing complexity, we analyze the statistical and structural properties of
the fluid-filled pore space that determines the flow statistics. We consider statistical network models based on regular lattices with the same statistical properties as the porous medium regarding coordination number and pore-size distribution. We consider regular lattices which are characterized by uniform coordination, and diluted lattices, and random lattices, which are characterized by a distribution of coordination numbers. Furthermore, we consider a detailed network model, which accounts for the spatial location of pores, their coordination numbers, and the sizes of pore bodies and throats. The flow behaviors estimated from these network models are compared to direct numerical single-phase flow simulations in the digitized images of a fully and partially saturated two-dimensional porous medium and different saturation degrees. We find that the statistical network models can capture the saturation dependence of permeability and tortuosity but are not able to reproduce velocity statistics of even the velocity range observed in the direct flow simulations. The detailed network models, in contrast, provide excellent estimates for all flow statistics. This indicates that the configuration and correlation of the fluid phase are crucial structural controls of the observed distribution of flow velocities.

\section*{Plain Language Summary}
Conceptualizing a porous media as a network of conductors sets a compromise between the oversimplifying conceptualization of the media as a bundle of capillary tubes and the computationally expensive and unobtainable detailed description of the media's geometry needed for direct numerical simulations. These models are abundantly being used to evaluate single and multiphase flow characteristics. The different flow characteristics are valuable in evaluating phenomena that may or may not be relevant for different applications. Here, we evaluate how different information about the pore space affects the ability of the network model to evaluate different flow characteristics. 

We found that the resistance of a media to the fluid flow can be estimated by the general stochastic features of the media (its size and connectivity). However, to account for more complex phenomena, such as solute transport and dispersion through the media, a piece of detailed information about the spatial location of the fluids is needed. 

\end{abstract}

\begin{highlights}
\item Flow through porous media can be conceptualized using Network models
\item In partially saturated media, the pore scale velocities distribution and tortuosity are governed by the phase configuration and not by pore geometry.
\item Network properties to match DNS flow statistics are identified
\end{highlights}

\end{frontmatter}


\section{Introduction}
Network models are commonly used to evaluate and analyze quasi-static \cite{blunt2001} and dynamic \cite{joekar2012}, single and multiphase flow through porous media. These models are used to predict the flow and transport through the media or to upscale pore-scale features to the Darcy scale hydraulic properties, such as retention and permeability.   
Conceptualization of the media as a pore network (rather than a bundle of capillaries) goes back to the work of \citet{Fatt1956}, where at its simplest form, it conceptualizes the media as a lattice of conductors, disregarding the pore shape. Another conceptualization involves an interconnected pore network with assumed circular pores \cite{Koplik1982}. A more complex (and realistic) modeled pore network is comprised of both bonds (relating to the pore throats) that represent most of the resistance and sites (relating to the pore bodies) that represent most of the capacitance \cite{Blunt2017}. In this context, many pore network models use a dual probability density function for the pore bodies' and throats' size distributions \cite{Dullien1975}. More detailed pore network models, accounting better for pores shapes and connections, such as equilateral triangles, squares, or stars \cite{Hoogland2016, Mason1991}, do not capture the full complexity of the actual pores but represent a compromise between the over-simplicity of a cylindrical capillary tube network and the directly-imaged pore shape models \cite{ben2023dynamics}.

Reduction in the water saturation degree decreases the volume fraction in which the flow can take place and the mean and size distribution of water-filled pores. The pore system connectivity is also reduced, which increases the structural tortuosity of the medium. In turn, the reduction of the saturation degree reduces the media's permeability and affects the flow velocity distribution. The velocity distribution broadens because of the formation of dead-end regions with slow velocities. These broad velocity distributions give rise to dispersive and non-Fickian solute transport \cite{VanGenuchten1979, ben2023solute, JM2017}.

The pore structure also affects the velocity distribution \cite{anna2017, Koplik1982}. The impact stems from the effect of the pore system geometry with wide pore bodies (low velocity) and narrow throats (high velocity) and the Newtonian fluid's viscosity with lower velocities near the grains surfaces and higher velocities in the pore throats centers. In this context, the capability of pore-networks conceptualization to evaluate the complex velocity fields needs to be investigated. Moreover, the importance of the different phenomena for different saturation degrees is unknown. Here, we use different network models based on different levels of information about the pore space and assumptions about the pore geometry to evaluate the importance of the different physical phenomena.

We first describe the experimental results of steady flow through variably-saturated milifluidic devices, the image analysis method used for pore segmentation and pore space characterization, and the numerical simulation of the flow on the exact pore space geometry. Second, we describe the different network models evaluated in this work (i.e., regular lattice, diluted lattice, random lattice, and detailed network representation of the pore system) and the different pore geometries under consideration (i.e., strip and parabolic). Lastly, we present the ability of the different models to evaluate the media's permeability, the spatial distribution of pressures and flow rates, the hydraulic tortuosity of the media, and the complex velocity distribution. The flow characteristics are compared to the values obtained by the numerical simulations. 

\section{Porous media and pore-scale flow}

In this section, we describe the porous media and pore-scale flow characteristics used in this work and to test the performance of the different network models given in Section~\ref{sec:Networks}.

\subsection{Pore space characterization} \label{sec:Exp}
\begin{figure}
 \includegraphics[width=.99\textwidth, center]{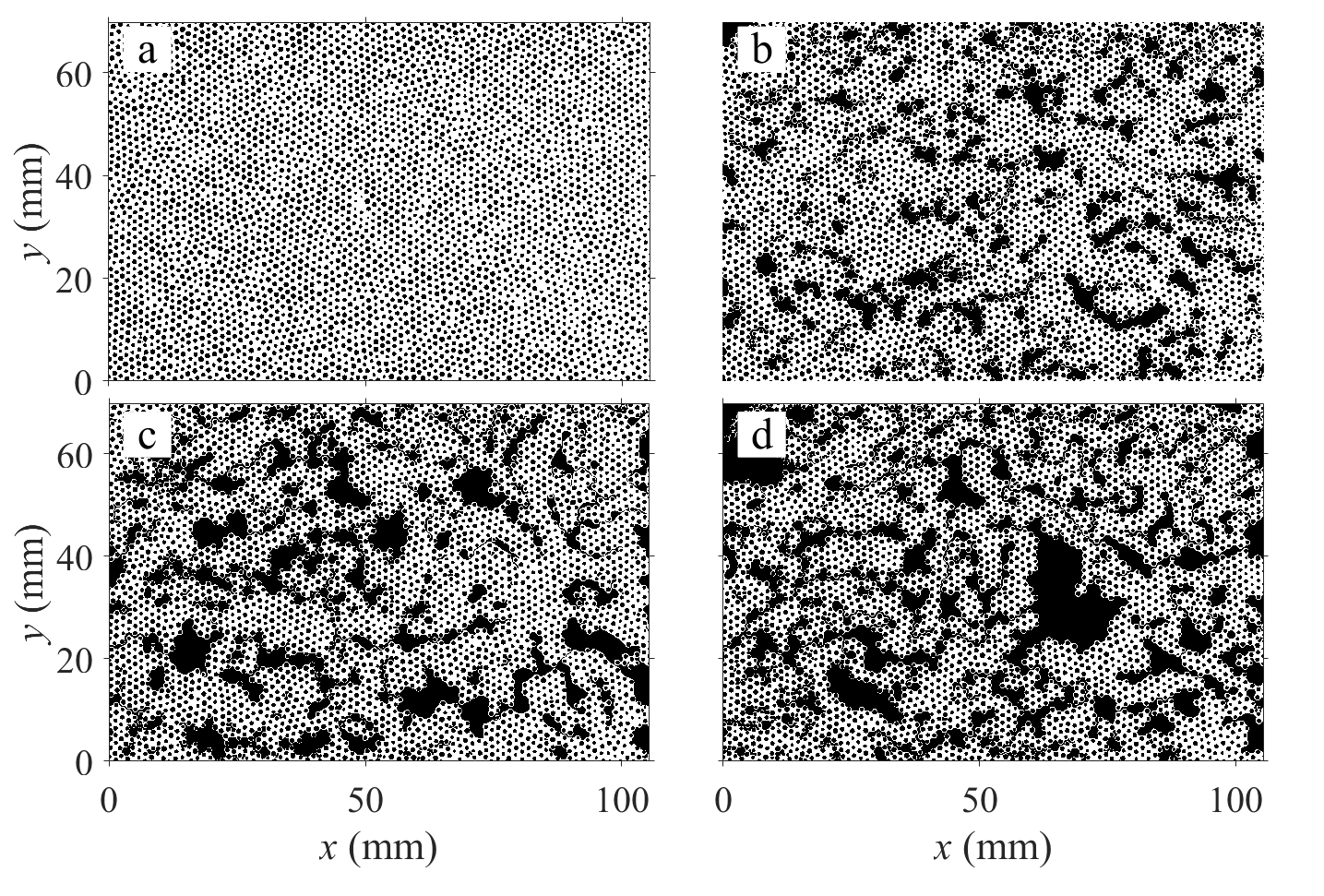}
 \caption{Binary images of the percolated flowing phase (in white) for a) saturated media ($S_e=1$) b) $S_e=0.80$, c) $S_e=0.73$, and d) $S_e=0.65$. Images were taken from the experimental work of \protect\citet{JM2017}. The solid medium comprises cylinders (pillars) with a mean diameter of $0.83$ mm (and standard deviation of $0.22$ mm), a height of $b=0.5$ mm, and a porosity of $0.72$. The pixel size of the 2D images is $\Delta x = 0.032$ mm.  The images do not include the entire original flow device and depict a flow domain of length $L=105$ mm ($3292$ pixels) along the main flowing axis ($x$-axis) and width $W=70$ mm ($2182$ pixels)\label{fig:media}}  
 \end{figure}

Porous media geometry is obtained from the binary images, i.e., the representation of the pores and the matrix as zeros and ones, of a milifluidic device at different saturation degrees (Figure~\ref{fig:media}) from the experimental work of \citet{JM2017}. The experiments established a steady flow of a water and glycerol mixture in variably saturated (i.e., containing also a stagnant air phase) hydrophilic millifluidic device. Four effective saturation degrees of $S_e=1.00$, $0.80$, $0.73$, and $0.65$ (Figure~\ref{fig:media}a,b,c, and d in accordance) are considered, where $S_e$ is defined as the fraction of the porosity ($\phi$) filled with the continuous and percolated liquid phase. Flow was controlled so that the phase distributions were not altered during the flow experiments, that is, steady-state imbibing conditions were attained. The flow rates, corresponding to a capillary number of about $3\cdot 10^{-5}$, are considered capillary-dominated.
\subsubsection{Image analysis and pore space partitioning} \label{sec:Image}
\begin{figure}
 \includegraphics[width=0.99\textwidth, center]{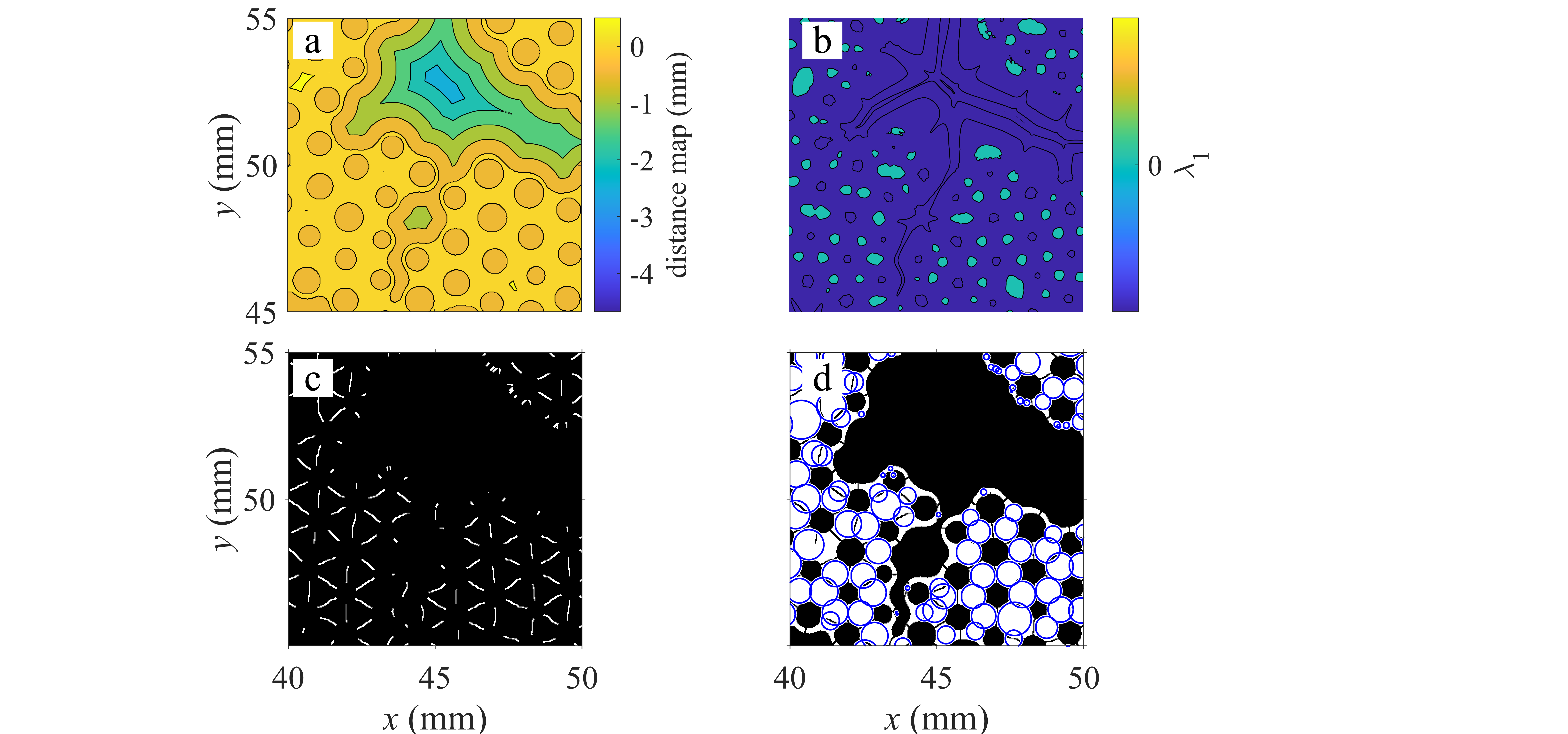}
 \caption{a) distance map b) small eigenvalues ($\lambda_1$) of the Hessian matrix map, c) binary image of the pore throats, and d) binary image of the segmented pore space in a zoomed-in section of the milifluidic device in the $S_e=0.65$ case. Blue circles in (d) present the pore body in terms of a maximal disc.}\label{fig:seg}  
 \end{figure}

The image analysis used in this work follows the methodology presented in \cite{ben2024Image}. In a nutshell, a ``$\lambda$-based'' medial axis method is used to segment the pore space into distinct pores. This method uses the binary images of the pore space (Figure~\ref{fig:media}) to form the distance map, i.e., the Euclidean distance of each point from the closest phases' interface (Figure~\ref{fig:seg}a). In the distance map, positive and negative values are assigned for the pore space and stagnant phase in accordance. Then, the distance map curvatures and the eigenvalues of the Hessian matrix of the distance map are calculated at each point (Fig.~\ref{fig:seg}b). A new binary image is then derived from the small eigenvalue ($\lambda_1$) map, in which negative values are assigned with 1 (i.e., the blue colored areas in Fig.~\ref{fig:seg}b) and the rest with 0. The union of the lines formed by the skeletonization of the new binary image and the complementary to the original binary image gives a binary image of the throats (Fig.~\ref{fig:seg}c). Thus, the union of the complementary to the throats image and the original binary image gives a new binary image of the segmented pores (Fig.~\ref{fig:seg}d). The pore bodies' sizes, in terms of the maximal disk that can fit within the pore, are given by the maximal value of the original distance map in each pore (blue circles in Fig.~\ref{fig:seg}d), i.e., not accounting for the throats for the confinement of the sphere. Similarly, the throats' sizes are the maximal values of the distance map on each distinct throat line, i.e., disregarding the tortuosity of the throat line. The eigenvalue ($\lambda_1$) map is processed using a Gaussian filter as discussed in the appendix of \citet{ben2024Image}. For the saturated case, a single-layer filter with a standard deviation ($\sigma=5$) is used. Conversely, a Gaussian pyramid filter is used for the partially saturated cases with a base level standard deviation ($\sigma_0=0.25$), standard deviation increments of ($\Delta\sigma=0.005$), and a scale space factor of ($\gamma=0.75$).     

\subsubsection{Saturation degree and pore space topology}

\begin{figure}
 \includegraphics[width=.99\textwidth, center]{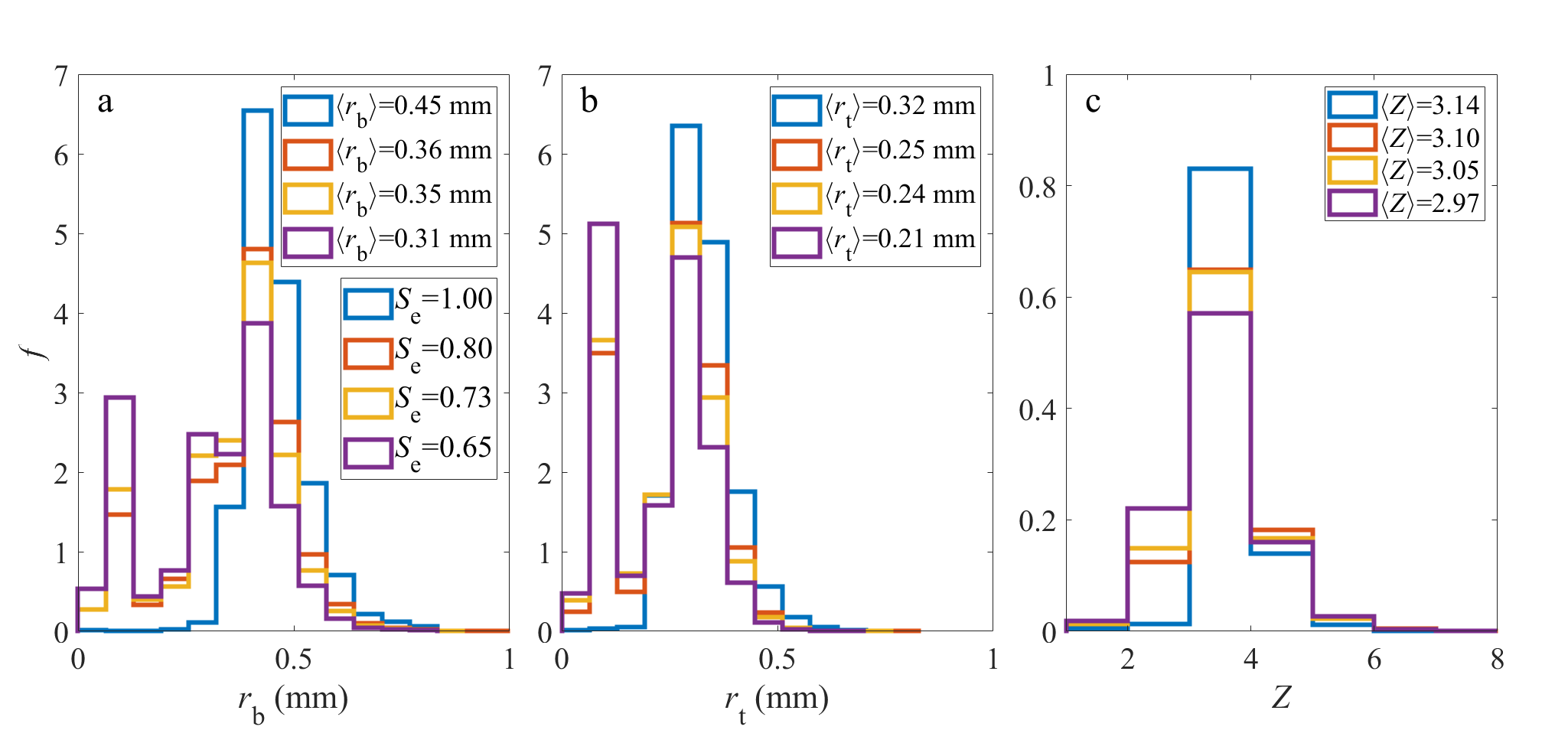}
 \caption{Probability density functions of the a) pore bodies radius ($r_t$), b) pore throats radius ($r_t$), and c) coordination number distributions of the different effective saturation degrees. The mean values are given in the figure labels. The pores bodies' and throats' sizes and connectivity are determined using the image analysis specified in section \ref{sec:Image}} \label{fig:psd}  
 \end{figure}

The effect of the saturation degree on the pore size distributions and
connectivity was discussed in \citet{ben2024Image}, and the effect of
$S_e$ on the velocity distribution in \citet{ben2023solute}, for
reader's convenience, we summarize it here. The reduction in $S_e$
reduces the mean water-filled pore size distribution because capillary
forces preferentially drive air to invade large pores. Moreover, air,
being the non-wetting phase, invades water-filled pores through their
center. Thus breaking up large pores into small water films
surrounding the grains. The latter mechanism forms a dual pore-size
distribution (Figure~\ref{fig:psd}a,b).
The throats' size distributions (Fig.~\ref{fig:psd}b) of this granular
media are of the same order of magnitude as the pore bodies, slightly
decreasing with the reduction in $S_e$.
The coordination number (Z, Fig.~\ref{fig:psd}c) is the number of
interconnected pores. The reduction in the saturation degree increases
the number of dead-end pores (Z=1) and of pores connected in series
($Z=2$). In turn, the mean coordination number ($\langle Z \rangle$)
reduces with the reduction in $S_e$, with a maximal value of $3.14$
and minimal $2.97$ for the saturated ($S_e=1$) and $S_e=0.65$ cases in
accordance.

 \subsection{Direct numerical simulations of pore-scale flow} \label{sec:Numerical}
 
The flow field is characterized in terms of its velocity distribution, stagnant region volume, permeability, and hydraulic tortuosity. The two-dimensional water flow field for each analyzed image is determined by solving the steady-state Stokes equation assuming an incompressible laminar flow
\begin{equation} \label{eq:stokes}
\mu \nabla^2 \bu-12 \mu \bu/b^2= \nabla p 
 \end{equation}
 \begin{equation*}
   \nabla \cdot \bu=0,
 \end{equation*}
 where $\bu$ [L/T] is the flow velocity, $\mu$ [ML$^{-1}$T$^{-1}$] the fluid's dynamic viscosity and $b$ [L] the device height. The term $-12 \mu \bu/b^2$ in (\ref{eq:stokes}) accounts for the viscous forces induced by the millifluidic device's top and bottom plates \cite{homsy1987}.

\subsubsection{Numerical setup}
 
We use the COMSOL \cite{COMSOL} multiphysics platform for the numerical simulation. A unit pressure gradient was imposed between the left (inlet) and right (outlet) boundaries, i.e., $p(0,y,t)=\rho g_cL$ and $p(L,y,t)=0$, with $\rho$ [ML$^{-3}$] the fluid density, $g_c$ [LT$^{-2}$] the gravitational constant, and L [L] the device length. The linearity of the flow equation (\ref{eq:stokes}) implies that both pressure and velocity (normalized by their mean values) are independent of the fluid properties (i.e., density and viscosity). A no-slip boundary condition was set at the phases' interfaces. Using the no-slip assumption on the liquid-gas interface during the liquid (wetting), phase flow was evaluated and justified in \citet{ben2023solute}.

\subsubsection{Flow statistics}

\paragraph*{Permeability} \qquad 

The permeability $k$ [L$^2$] of a medium is defined as the flux of a viscosity unit that a unit pressure (P, [ML$^{-1}$T$^{-2}$]) gradient will invoke \cite{Muskat1931}
\begin{equation} \label{eq:permeability}
k=\frac{q_m\mu}{|\nabla P|}=\frac{q_m\mu}{\rho g_c}.
\end{equation}
The flux $q_m$ is given in terms of the volumetric flow rate, or discharge rate $Q$ as $q_m=\frac{Q}{Wb}$ [L T$^{-1}$], which can be measured at any medium cross-section. 

\paragraph*{Distribution of flow speeds} \qquad 

We determine the probability density function (PDF) of Eulerian flow speeds $u=|\bu|$, denoted by $f(u)$, through areal sampling from the direct numerical flow simulations as
\begin{align}
\label{eq:fu}
f_u(u_k) = \frac{\sum_i \mathbb I(u_i,u_k) A_i}{\Delta u_k\sum_i A_i},
\end{align}
where $A_i$ [L$^2$] is the area of the $i$-th mesh element, and $u_i$ is the flow speed at the Gaussian point of the $i$-th mesh element. The indicator function $\mathbb I$ is one if the velocity $u_i$ is within the interval $(u_k, u_k + \Delta u_k]$ and zero otherwise.

\paragraph*{Hydraulic tortuosity} \qquad 

The hydraulic tortuosity $\chi$, which measures the ratio of averaged trajectory length to the linear distance, accounts for the divergence of the streamline from the macroscopic flow direction. Thus, it can be calculated by the ratio of the mean flow speed $\overline u_e$ [LT$^-1$] and mean flow velocity (in the flow direction) $\overline u = q_m/S_e\phi$~\cite{koponen1996}
\begin{equation} \label{eq:chi}
\chi = \frac{\overline u_e}{\overline u}.
\end{equation}

\paragraph*{Stagnant regions} \qquad 


The media permeability, hydraulic tortuosity, 
and velocity distribution of the different
saturation degrees of the numerical simulation are considered the
``ground truth'' and the basis for estimating the applicability of the
different network models.

\section{Pore network models} \label{sec:Networks}
We present here pore network models of different complexity to represent the
porous media discussed in the previous section. We first summarize the basic
methodology for solving flow on networks. Then, we discuss two representations of the
pore geometry and the corresponding pore-scale flow distribution and target flow
statistics. With these preparations, we define the network models used to
represent the flow properties in the model porous media presented in the previous section. 

Pore network models represent a porous medium as a network of sites and bonds, which connect them. Sites are typically located at the centers of pore bodies, the bonds align with the pore throats. Both are identified during the pore-space partitioning process. These basic network features can be supplemented with information on the pore shape, diameter, and other relevant geometric and hydraulic characteristics. 
In the following, we use the terminology "sites" and "bonds" when we refer to the network model and "body" and "throat" when we refer to the actual porous medium.  

\subsection{Flow on networks}
The flow rate $Q_{ij}$ along the bond connecting the sites $i$ and $j$ is given by
\begin{align}
\label{eq:k1}
Q_{ij} = - g_{ij} (P_j - P_i),
\end{align}
where  $P_i$ and $P_j$ are the pressures at sites $i$ and $j$. The
bond conductance is $g_{ij} = \kappa_{ij}/\ell_{ij}$ [M$^{-1}$L$^4$M$^{-1}$] with $\ell_{ij}$ [L] the
bond length and $\kappa_{ij}$ [M$^{-1}$L$^3$M$^{-1}$] the bond conductivity. The bond conductivity
depends on the pore geometry as discussed 
in detail in Subsection~\ref{sec:pore_geo}.  Note that $Q_{ij} = - Q_{ji}$. If $Q_{ij} > 0$, flow
is directed from node $i$ to $j$ and vice versa. Volume conservation
requires that the sum of the flow rates in the bonds meeting at a site
$j$ is zero,
\begin{equation} \label{eq:k2}
\sum_{[ji]} Q_{ij} = 0,
\end{equation}
where the notation $\sum_{[ji]}$ denotes summation over the sites $j$
that are connected to site $i$. The coordination number of site $i$, that is, the number of
sites $j$ connected to site $i$ is $Z_i = \sum_{[ji]}$. The flow rate $\mathcal
Q_i$ at an internal site $i$ of the network is
evaluated from the sum of the flow rates $Q_{ij}$ through the bonds
that are connected to the site $i$ as
\begin{equation} \label{eq:Q_s}
\mathcal Q_{i}= \frac{1}{2}\sum_{[ji]} |Q_{ij}|.
\end{equation}
In words, $\mathcal Q_i$ equals half the sum of the absolute values in the rows of the flow rate matrix $\mathbf Q$. 
Combining equations~\eqref{eq:k1} and~\eqref{eq:k2} gives the following system of
equations at all internal sites of the network,
\begin{equation} \label{eq:matrix_Ohm}
\mathbf L \cdot \mathbf P = \mathbf 0,
\end{equation}
where $\mathbf P$ is the vector of pressures at all sites of
the network, $\mathbf L = \mathbf D - \mathbf G$ is the weighted Laplacian matrix.
It is given by the difference of the degree matrix $\mathbf D$ and the conductance matrix $\mathbf G$.
The degree matrix is a diagonal matrix that contains the degrees of each node,
that is $D_{ik} = \delta_{ik} \sum_{[ji]} g_{ij}$. The weight matrix contains the
conductances between connected sites $G_{ij} = g_{ij}$ and $0$ else. The
pressure boundary conditions used here are implemented by defining
$\mathbf L'$ such that $L'_{ij} = L_{ij}$ for all sites $i$ that are not,
and $L'_{ij} = \delta_{ij}$ for all sites $i$ that are in the inlet or
outlet boundaries such that
\begin{equation} \label{eq:matrix_Ohm2}
\mathbf L^\prime \cdot \mathbf P = \mathbf b,
\end{equation}
where the vector $\mathbf b$ on the
right side of Eq.~\eqref{eq:matrix_Ohm2} contains the boundary conditions. We assign the pressure 
$b_{j} = p_0 = \rho g_c L$ at the sites $j$ at the inlet boundary and $b_j = 0$ else.  

\subsection{Pore geometries and conductivities}\label{sec:pore_geo}
Pore network models can also include representations of the geometry
of the pore space \cite{Koplik1982,gostick2016openpnm,Liu2024}, in which the
bodies represent most of the phase capacitance and throats most of the
resistance to the flow \cite{Blunt2017}. 


\subsubsection{Rectangular pore shape}
The rectangular ("strip-work") pore shape is
characterized by the pore channel height $b$ and width $w$.
Recall that from the pore-space partitioning, we know the sizes of the pore bodies
and pore throats and their distributions (see Fig.~\ref{fig:psd}). Since most of
the resistance to the flow is in the throats, we set $w = 2 r_t$ (black line in Figure~\ref{fig:gpdf}b).
For a rectangular channel, the conductivity is given by \cite{Boussinesq1868}
\begin{equation} \label{eq:Boussinesq}
\kappa = \frac{b^3 w}{12 \mu}-\frac{16
  b^4}{\mu}\sum_{n=1}^{\infty}\frac{\cosh(\frac{2\beta_nw}{b})-1}{\beta_n^5
  \sinh(\frac{2\beta_nw}{b})},
\qquad \beta_n=(2n-1)\pi,
\end{equation}
The distribution of bond conductivities is then given by 
\begin{align}
\label{pk_rect}
p_\kappa(k) = \int\limits_0^\infty d r f_t(r) \delta[k - \kappa(r)].
\end{align}

Note that the pore throat, in principle, has zero length as it marks the boundary
between two pores. Thus, assigning a finite length to the bond underestimates
the conductance of the pore. The velocity profile in a rectangular pore is
approximated by the Hagen-Poiseuille equation
\begin{align}
\label{eq:parabolic_u}
u(y) = \frac{3 Q}{2b w} \left(1 - \frac{y^2}{w^2}\right),
\end{align}
where the pore flow rate is $Q = \kappa |\Delta P|/\ell$, where $\Delta P$ is the
pressure drop along the length $\ell$ of the pore. 
\subsubsection{Parabolic pore shape \label{sec:parabolic}}

\begin{figure}
 \includegraphics[width=0.9\textwidth]{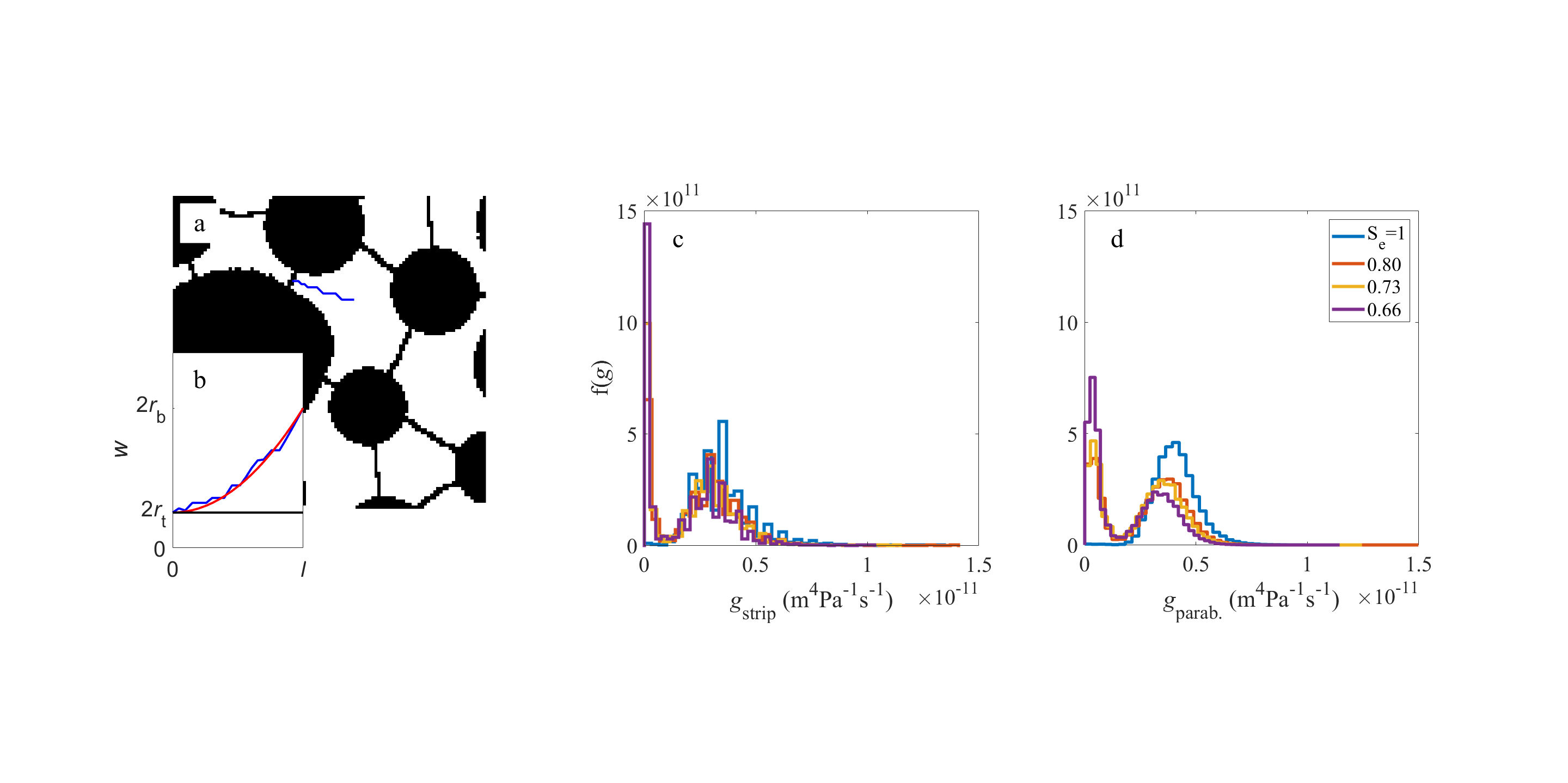}
 \caption{a) A zoomed-in section in the $S_e=0.65$ case, the blue line depicts the bond medial-axis ($\mathbb{x}$) from the throat to the pore centers, b) the bond aperture evaluated from twice the distance map (blue) value (i.e., twice the distance from the nearest solid-pore interface), the strip geometry assumption (black), and the parabolic geometry (red).  
 And bonds conductance probability density function ($f(g)$) for c) strip
   (Eq.~\ref{pk_rect}) and d) parabolic pore geometries
   (Eq.~\ref{pk_parabol}).}\label{fig:gpdf}
 \end{figure}

The second geometry under consideration assumes a parabolic shape of the pore (red line in Figure~\ref{fig:gpdf}b). 
This parametrization of the aperture is agile enough to give a good approximation to the actual aperture, for example, 
given by the minimal distance from the pore-media interface of the medial axis (blue line in Figure~\ref{fig:gpdf}a and b). 
where the aperture ($w$) of a pore changes from approximately the pore throat diameter ($2r_t$) at one end ($\mathbb{x}=0$) to the pore body diameter
($2r_b$) at its
other end ($\mathbb{x}=l$), where $\mathbb{x}$ is the axis of the main flow
direction taken to be perpendicular to the pore throat facing away
from the pore center. Moreover, the parabola has a minimal value at the pore
throat, which means that the derivative of $w(x)$ at
the throat location  $\mathbb{x}=l$ is zero. Then, the pore aperture is given by
\begin{equation} \label{eq:aperture}
w(\mathbb{x}) = 2r_t+2(r_b-r_t)\frac{\mathbb{x}^2}{l^2}
\end{equation}
%

For laminar flow, the effective pore conductance $\kappa_e$ is evaluated by the
harmonic sum of the local conductivities along the pore channel with variable
width $w(x)$ by using Eq.~\ref{eq:Boussinesq} for $\kappa(\mathbb x)$
\begin{equation} \label{eq:parabolic}
\frac{1}{\kappa_e} = \frac{1}{\ell} \int\limits_0^\ell \frac{dx}{\kappa(x)}.
\end{equation}
Note that $\kappa_e = \kappa_e(r_b,r_t)$ is a function of the body and throat
radii because $\kappa(x)$ depends on $w(x)$. Assuming that $r_t$ and $r_b$ are
independent variables, that is, disregarding the correlation between the bodies
and the throats and also disregarding the correlation between $r_b$ and the
coordination number ($Z$), the distribution of
conductivity can be determined as
\begin{align}
\label{pk_parabol}
p_\kappa(k) = \int\limits_0^\infty d r \int\limits_0^{r} d r' f_b(r) f_t(r')
\delta[k - \kappa_e(r,r')].
\end{align}
Note that we use the condition that $r_t < r_b$. This condition
causes a bias towards smaller throats and larger bodies. The magnitude of this
bias depends on the overlap of the pore body and throat size
distributions (Figure~\ref{fig:psd}). For a small overlap, the bias is
small. On the other hand, a significant overlap means that the pore bodies
and throats are of the same size. In this case, the pore geometry is
essentially rectangular. In the model media, the rectangular pore geometry
(Fig.~\ref{fig:gpdf}c) has many more small conductors for
the unsaturated cases than the parabolic geometry (Fig.~\ref{fig:gpdf}d). The
saturated cases (blue lines) are very similar in both magnitude and distribution
because the differences between $f(r_b)$ and $f(r_t)$ are not very large
(Fig.~\ref{fig:psd}).

\begin{figure}
 \includegraphics[width=0.9\textwidth]{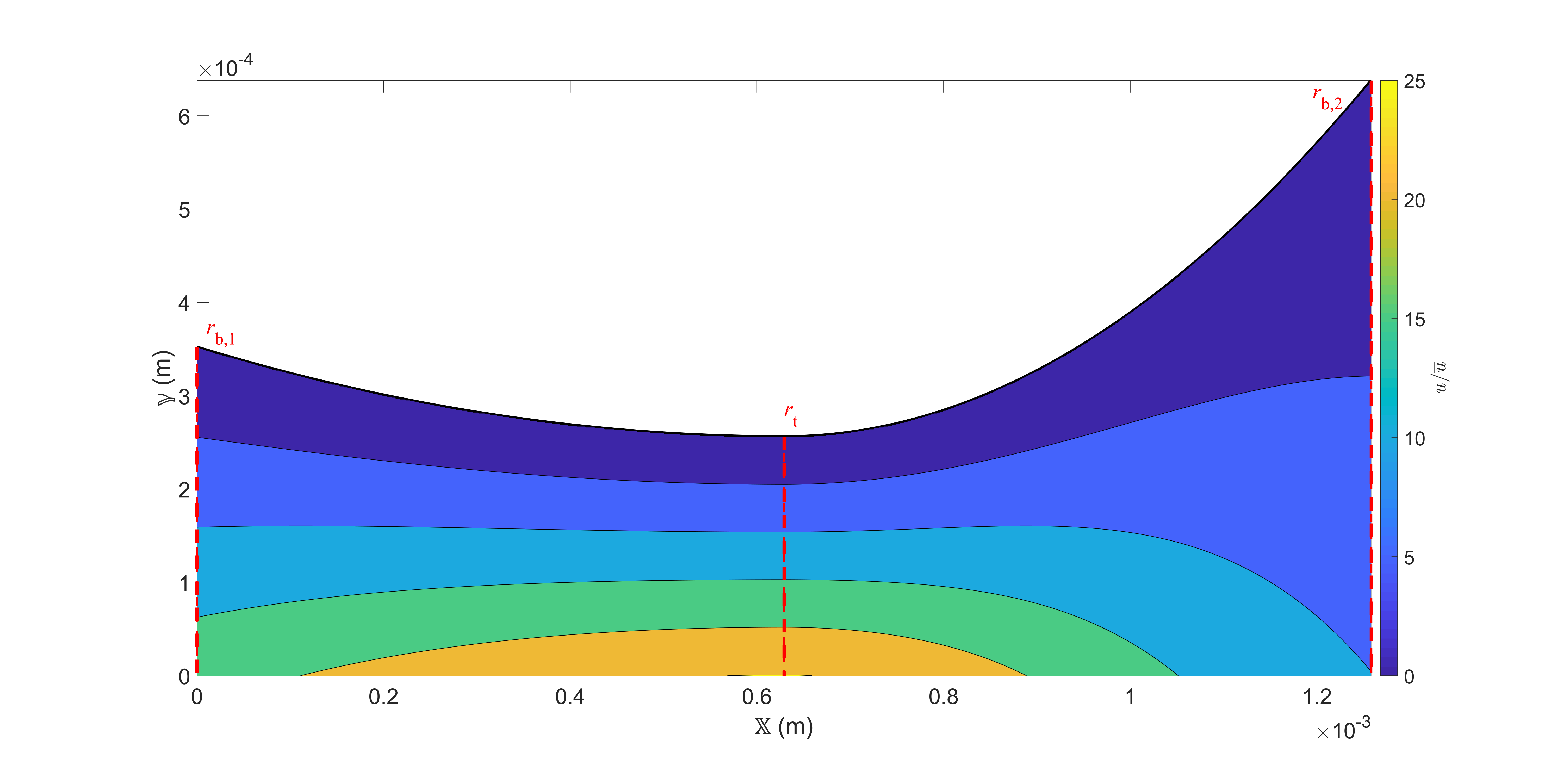}
 \caption{The velocity distribution in a single bond in the detailed
   network. The axis $\mathbb{y}=0$ is the center of the bond, which is assumed
   symmetrical around this axis. 
 \label{fig:intrabond_u}}  
 \end{figure}

For a more detailed representation, the pore space is reconstructed by joining semi-parabolas at the pore
throat such that they have the same $r_t$, but in general, different
$r_b$, see Figure~\ref{fig:intrabond_u}. 
The pore geometry also affects the intra-pore velocity distribution with higher
velocity in the narrow section of the pore throat due to volume
conservation. That is, the mean cross-sectional velocity is inversely
proportional to the $w(x)$. The no-slip condition at the pore-grain boundary
gives a parabolic velocity profile for laminar flow conditions. Here, for
simplicity, we approximate the velocity profile across the pore by the
Hagen-Poiseuille equation
\begin{equation} \label{eq:u_distribution}
u(\mathbb{x},\mathbb{y})=\frac{3Q}{2 b w(\mathbb{x})}
\left(1-\frac{\mathbb{y}^2}{w(\mathbb{x})^2}\right),
\end{equation}
where $Q = \kappa_e |\Delta p|/\ell$ is the pore flow rate. Note that we set
the average cross-sectional velocity to $\overline{u_\mathbb{x}} =
{Q_b}/{bw(\mathbb{x})}$. The maximum flow velocity is $u_m=3
\overline{u_\mathbb{x}}/2$ at the center of the conduit.  The
resulting velocity distribution is illustrated in Figure~\ref{fig:intrabond_u}.

\subsection{Flow statistics}

\paragraph*{Permeability} \qquad 
The permeability $k$ of the network is
determined from the flow rates at the inlet sites, which are given by
$\mathcal Q_j = \sum_{[ji]} Q_{ji}$, where $j$ is a site at the inlet boundary. The
flux $q_m$ is then given by $q_m = \sum_{j} Q_{j}/(Wb)$. From $q_m$ we
obtain the network permeability according to the expression in
Eq.~\ref{eq:permeability}. For regular and diluted networks, $k$ can be 
evaluated using the Effective Medium Approximation (EMA), see \ref{sec:EMA}.  

\paragraph*{Distribution of flow speeds} \qquad
The distribution of flow speeds in a
single bond is denoted by $p_{ij}(u)$. For the parabolic
profile~\eqref{eq:parabolic_u} it is given by 
\begin{align}
p_{ij}(u) = \frac{b w_{ij}}{3 Q_{ij}\sqrt{1 - 2 u b w_{ij}/Q_{ij}}}
\end{align}
for $0< u < 3 Q_{ij}/2 bw$. For the velocity profile~\eqref{eq:u_distribution} it is
accordingly given by
\begin{align}
p_{ij}(u) = \frac{1}{\ell_{ij}}\int\limits_{0}^{\ell_{ij}} dx \frac{b w_{ij}(x)}{3
  Q_{ij}\sqrt{1 - 2 u b w_{ij}(x)/3 Q_{ij}}}. 
\end{align}
The speed distribution across the network is then obtained by the weighted
sum of the $p_{ij}$, that is, 
\begin{align}
f_u(u) = \frac{\sum_{i>j} p_{ij}(u) A_{ij}}{\sum_{i>j} A_{ij}}. 
\end{align}
%
\paragraph*{Distribution of flow rates} \qquad
The distribution of flow rates $Q_{ij}$
is determined in analogy to the distribution of flow speeds as
\begin{align}
f_Q(Q_k) = \frac{\sum_{i>j} \mathbb I(Q_{ij},Q_k)}{\Delta Q_k}. 
\end{align}

\paragraph*{Hydraulic tortuosity} \qquad 
We suggest to estimate the hydraulic tortuosity of the network
as the ratio between the average flow rate magnitude and the
average flow rate in the network, that is,
%
\begin{equation} \label{eq:tau}
\tau = \frac{\sum_{i>j} |Q_{ij}|}{\sum_{i>j} Q_{ij} \cos(\alpha_{ij})}, 
\end{equation}
where $\alpha_{ij}$ is the bond angle with the $x$-axis. We can define an average bond velocity as
\begin{align}
v_{ij} = \frac{Q_{ij}}{A_{ij}},
\end{align}
where $A_{ij}$ is the average surface area of the bond cross-section. If we
assume that the bond lengths are not very variable and can be replaced by the
network average, $\ell_{ij} = \langle \ell \rangle$, definition~\eqref{eq:tau}
becomes
\begin{align}
\tau = \frac{\sum_{i>j} |v_{ij}| V_{ij}}{\sum_{i>j} v_{ij} \cos(\alpha_{ij}) V_{ij}} = \frac{\langle v \rangle}{\langle v_x \rangle} = \chi, 
\end{align}
where we set the bond volume $V_{ij} = A_{ij} \langle \ell \rangle$.

\subsection{Network models}
 
\subsubsection{Regular networks} \label{sec:regular}

\begin{wrapfigure}{r}{0.25\textwidth}
\includegraphics[width=0.9\linewidth]{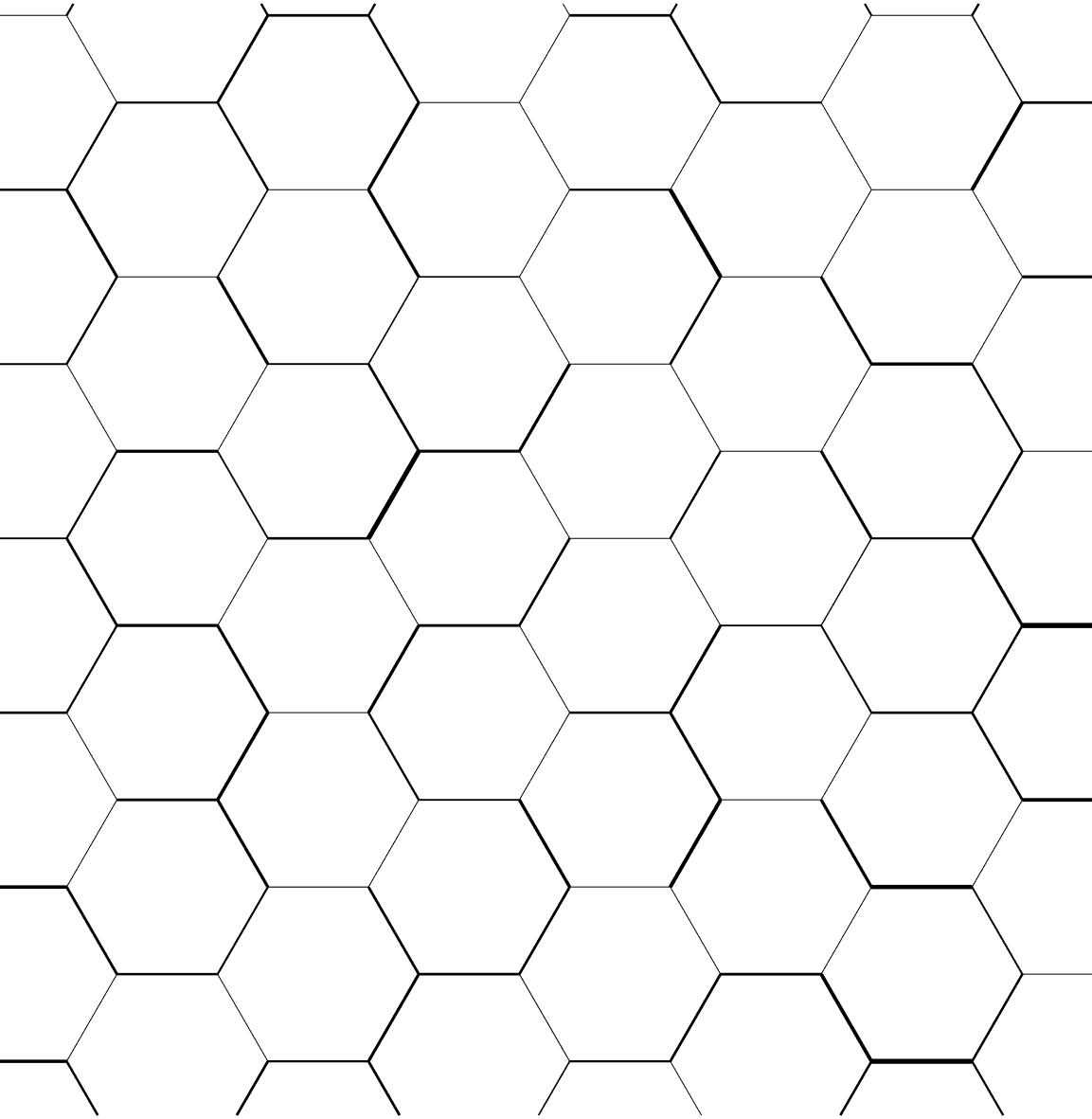} 
 \caption{Regular (hexagonal) network.}
\label{fig:reg_net}  
\end{wrapfigure}

A regular network (Figure~\ref{fig:reg_net}) refers to a lattice with a constant
coordination number ($Z$). Regular networks are the most basic networks under
consideration in this study. They require only a pore size distribution as input
parameters and an estimate of the mean coordination number to decide the
lattice geometry. The coordination number is considered the single most
important topological feature of the medium \cite{hunt2017}. It has been shown
that regular networks and topologically disordered networks variable $Z$ have
similar transport properties if they share the same mean coordination number
$\langle Z \rangle$ \cite{Sahimi1990}. The mean coordination number of the study
media presented in Section~\ref{sec:Exp} is close to $\langle Z \rangle = 3$ for
all cases, see Fig.~\ref{fig:psd}c. Therefore, a regular hexagonal (honeycomb)
lattice with $Z=3$ and a constant bond length $\ell$ is used. Moreover,
for each saturation degree, the number of sites of the corresponding honeycomb
network model is equal to the number of pores obtained from the pore space
segmentation in Subsection \ref{sec:Image}, while the physical dimensions of the
network are kept constant. This implies that the bond lengths of the different
network models are slightly different. The longest diagonal of the network is
aligned with the main flow direction. The bond conductances are assigned
randomly from the distribution $f(g)$ (Fig.~\ref{fig:gpdf}). As discussed in
Subsection~\ref{sec:pore_geo}, we consider both rectangular and parabolic pore geometries. This process disregards any spatial correlation
in the pore size distribution, which is in line with the pore structure of the porous media considered 
in this study. For correlated networks, we refer readers to
\citet{Sahimi1998}. 

\subsubsection{Diluted regular networks} \label{sec:diluted}

\begin{wrapfigure}{r}{0.25\textwidth}
\includegraphics[width=0.9\linewidth]{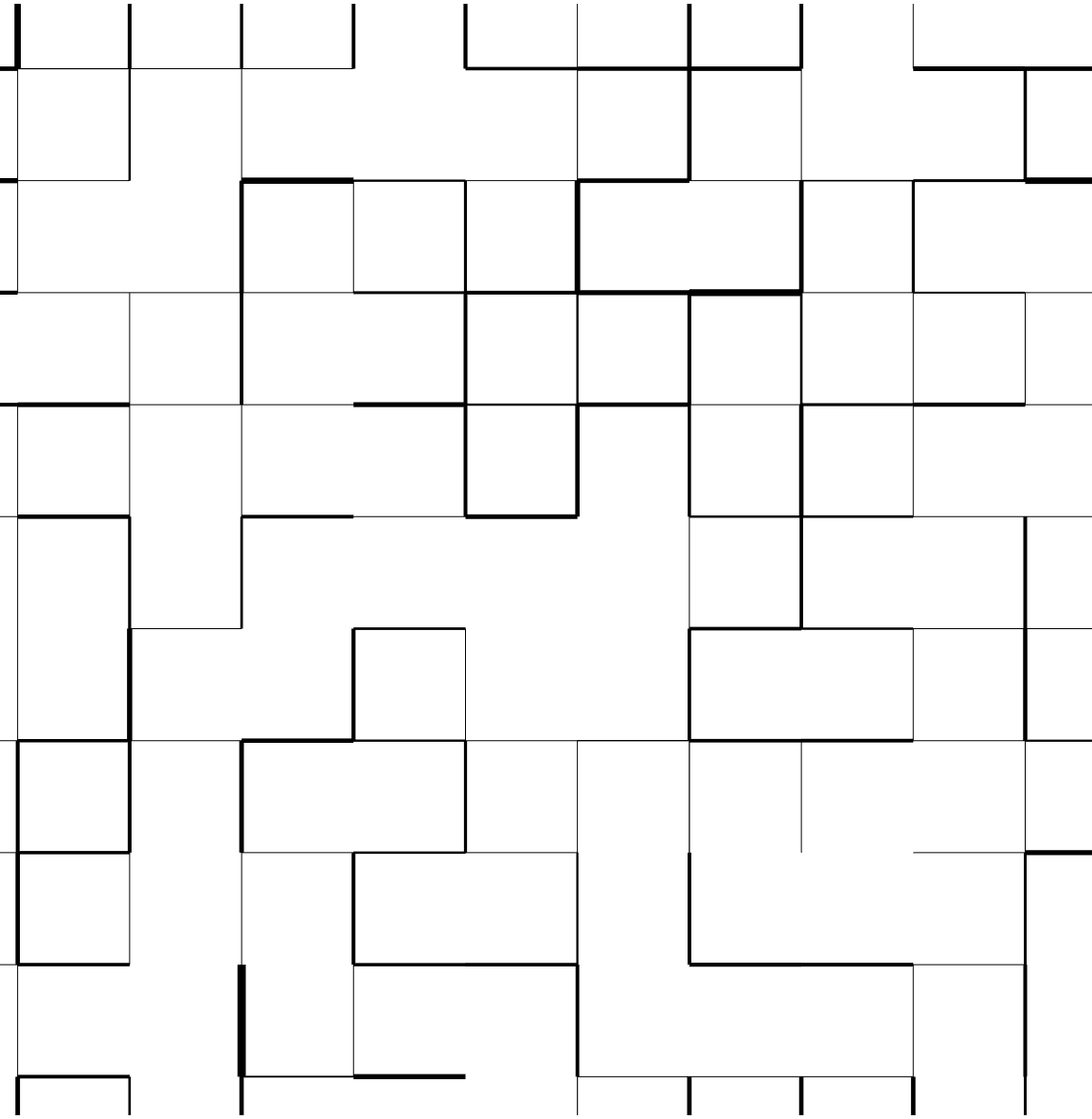} 
 \caption{Diluted regular network.
\label{fig:Sq_net}}  
\end{wrapfigure}

The diluted regular networks comprise a more detailed variation of the regular
networks (Figure~\ref{fig:Sq_net}). In these networks, a regular lattice with
$Z_{lattice} > \langle Z \rangle$ is imposed. Then
bonds are eliminated until the mean coordination number of the diluted network is equal to the mean coordination number of
the porous medium, $\langle Z_{diluted} \rangle = \langle Z \rangle$.
That is, the coordination number is variable but does not necessarily follow the distribution of coordination
numbers in the porous medium. As for the regular networks, the bond conductances are assigned
randomly from the distribution $f(g)$ (Fig.~\ref{fig:gpdf}) for both
rectangular and parabolic pore geometries. For each medium, the mean coordination number can be evaluated
from the number of bonds ($N_b$) and sites ($N_s$) through the relation $\langle
Z \rangle=2 {N_b}/{N_s}$. It should be noted that for irregular $3D$ media,
this ratio does not necessarily hold because a throat might connect more than
two pores \cite{ben2024Image}. 
The regular and diluted models
 (square and hexagonal) has the same number of bonds ($N_b$) equal to
 the number of pore throats taken from the image analysis. Since we
 keep the original device dimension, this means that the bond lengths are slightly different ($l_{sq}>l_{hex}$).

For the diluted regular networks, we use a regular square lattice ($Z_{grid}=4$) for all cases ($S_e$) and a hexagonal lattice
($Z_{grid}=3$) only for the lowest saturation degree ($S_e= 0.65$ for which the mean coordination number is $\langle Z_{media} \rangle=2.97$). Comparing this case with different lattices allows us to separately
evaluate the role of the lattice dimensionality and dilution.
We start with a fully saturated grid, i.e., with a number of bonds equal to the
number of throats obtained from the pore segmentation using the image
analysis. Then, the conductors are randomly assigned from the distribution
$f(g)$. As in the regular network model, we compare the strip and
parabolic pore geometries (described in Subsection \ref{sec:pore_geo}) by
accounting for the different conductances support ($g$) and distribution
($f(g)$). Then, the grid is diluted to obtain the $\langle Z \rangle$ of each
case (of different $S_e$), and in consequence, $N_s$ equals the number of pores
in the device. The dilution method applied here follows a simplification to the
protocol described by \cite{Raoof2010}. Each bond is randomly assigned with a
dilution (or elimination) number. Then, an elimination threshold value is used
so that all bonds with an elimination number larger than the threshold are
assigned zero conductivity. The elimination threshold is related to the mean
coordination number through
\begin{equation} \label{eq:dilution_th}
f_{et}=\langle Z_{\text{media}} \rangle /Z_{\text{grid}},
\end{equation}
where  $Z_{\text{grid}}$ is the coordination number of the regular lattice
before dilution. The conductance distribution is adjusted to account for the dilution, i.e.,
adding $g=0$ with a dilution probability of $p(g=0)=1-N_b/N_s$. 

The random elimination of bonds by assigning zero conductivity $k=0$ can
create singular points or regions, that is, subregions of the network that are not
connected to the inlet and outlet. Such regions are artifacts of the network generation
procedure and do not reflect the real medium structure. From a numerical point of
view, disconnected regions may cause numerical problems since they
lead to a singular or ill-conditioned Laplacian matrix. The challenge of addressing
this singularity issue was raised in several diluted network models \cite{Raoof2010}. Usually, this issue is solved by preconditioning the matrix
or by using a search algorithm to find and remove all ill-connected sites and
clusters. In this contribution, we use the Singular Value
Decomposition (SVD) method as an efficient way to eliminate the singular points of
the matrix \cite{Ben2024SVD}. Regarding the flow behavior, diluted regular networks may display more
stagnant regions than regular networks, that is, sites of very low or zero velocity
due to the occurrence, for example, of dead-end pores. 

\subsubsection{Random network} \label{sec:random}

\begin{wrapfigure}{r}{0.25\textwidth}
\includegraphics[width=0.9\linewidth]{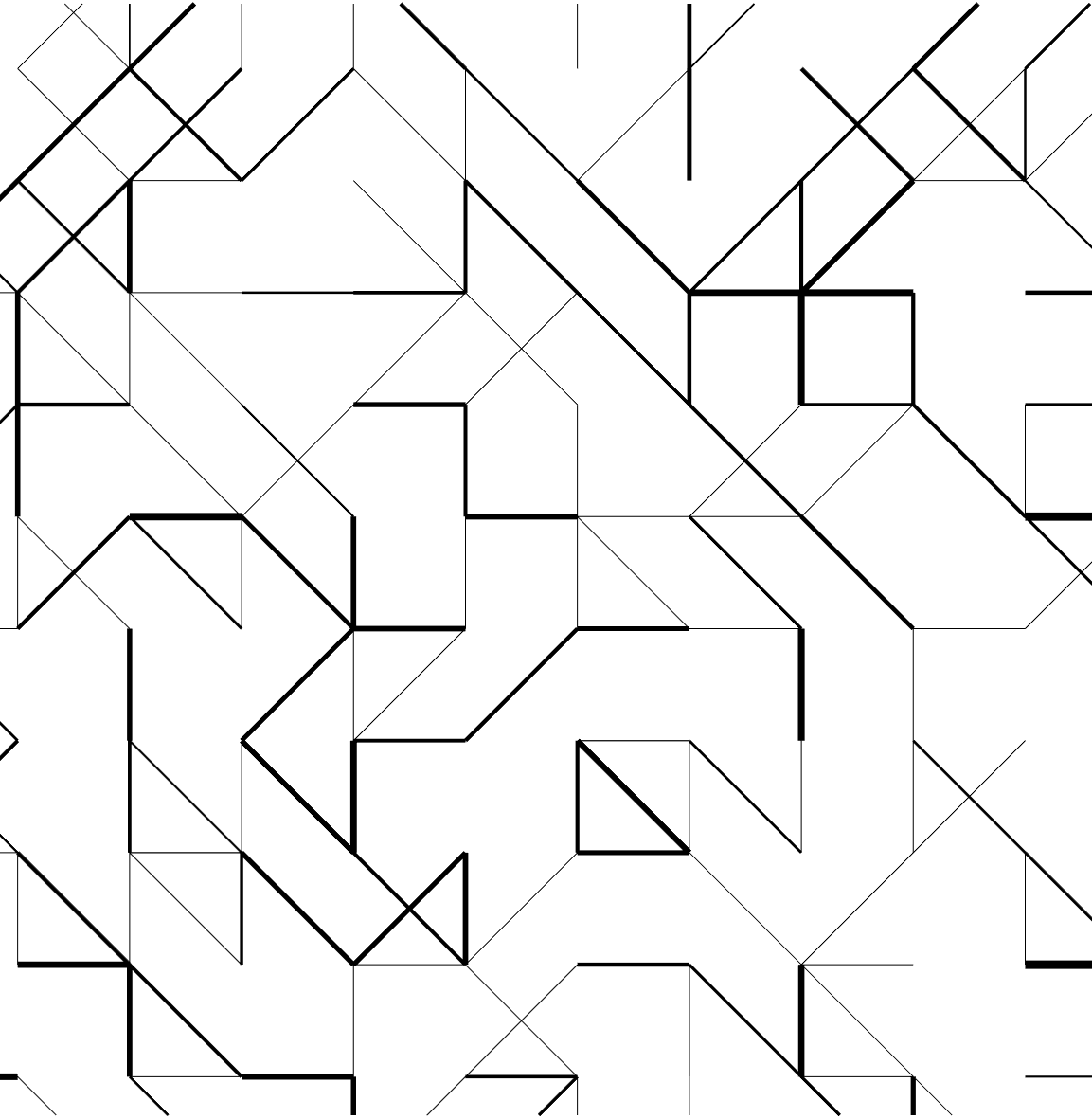} 
 \caption{Random network.} \label{fig:Oct_net}
\end{wrapfigure}

In this section, we describe the generation of a random network
(Figure~\ref{fig:Oct_net}) that has the same coordination number
distribution as the porous medium (not just the same mean). We start from a regular square
lattice of sites that are also connected along the diagonals, i.e.,
with a constant coordination number of $Z_{lattice} = 8$. This value
is equal to the largest observed coordination number ($Z=8$,
Fig.~\ref{fig:psd}c). Unlike for the previous
networks, in this case, the length is not constant because the
diagonals are longer than the horizontal and vertical bonds by a
factor of $\sqrt{2}$. As for the previous networks (sections \ref{sec:regular} and
\ref{sec:diluted}), conductance values are assigned randomly to all the bonds
from the distribution $f(g)$. Adjusting the distribution of 
coordination numbers of the lattice to a prescribed distribution is done
iteratively through diluting and concentrating schemes as described in
the following. 

The algorithm is initialized by assigning a target coordination number
$\mathbb Z_j \leq Z_{lattice}$ to each site $j$ drawn randomly from the
distribution $f(Z)$ (Fig.~\ref{fig:psd}c), and by assigning a random elimination
number to each bond of a site from a uniform
distribution between $0$ and $1$. All site coordination numbers at initialization are  $Z_j^{(0)} \geq Z_{lattice}$. At the next step (dilution step), if
a site coordination number is larger than the target value ($Z_j^{(0)} > \mathbb Z_j$), the bond with the smallest elimination number is removed by assigning zero conductance. 
After the dilution step, all coordination numbers are updated. The dilution step is followed by a concentration step. If the grid coordination number at the site is smaller than the target value ($Z_j^{(1)} < \mathbb Z_j$), then a random bond is added (to one of the available locations, keeping the initial conductance value) and a new
random elimination number is assigned to the bond. After the concentration step, all coordination numbers are updated, and a dilution step follows. These
  steps are repeated until the distribution of coordination numbers in
  the random network has converged to the target distribution. As a convergence criterion, we use a threshold value for the difference between the target and lattice mean coordination number. That is, the algorithm has converged after the $k$th iteration if $|\langle \mathbb Z \rangle - \langle Z^{(k)} \rangle |<10^{-4}$. 
For the media under consideration, the coordination number distribution converges to the
prescribed distribution after less than 30 iteration steps. 

\subsubsection{Detailed network representation of the pore system} \label{sec:detailed}

\begin{wrapfigure}{r}{0.25\textwidth}
\includegraphics[width=0.9\linewidth]{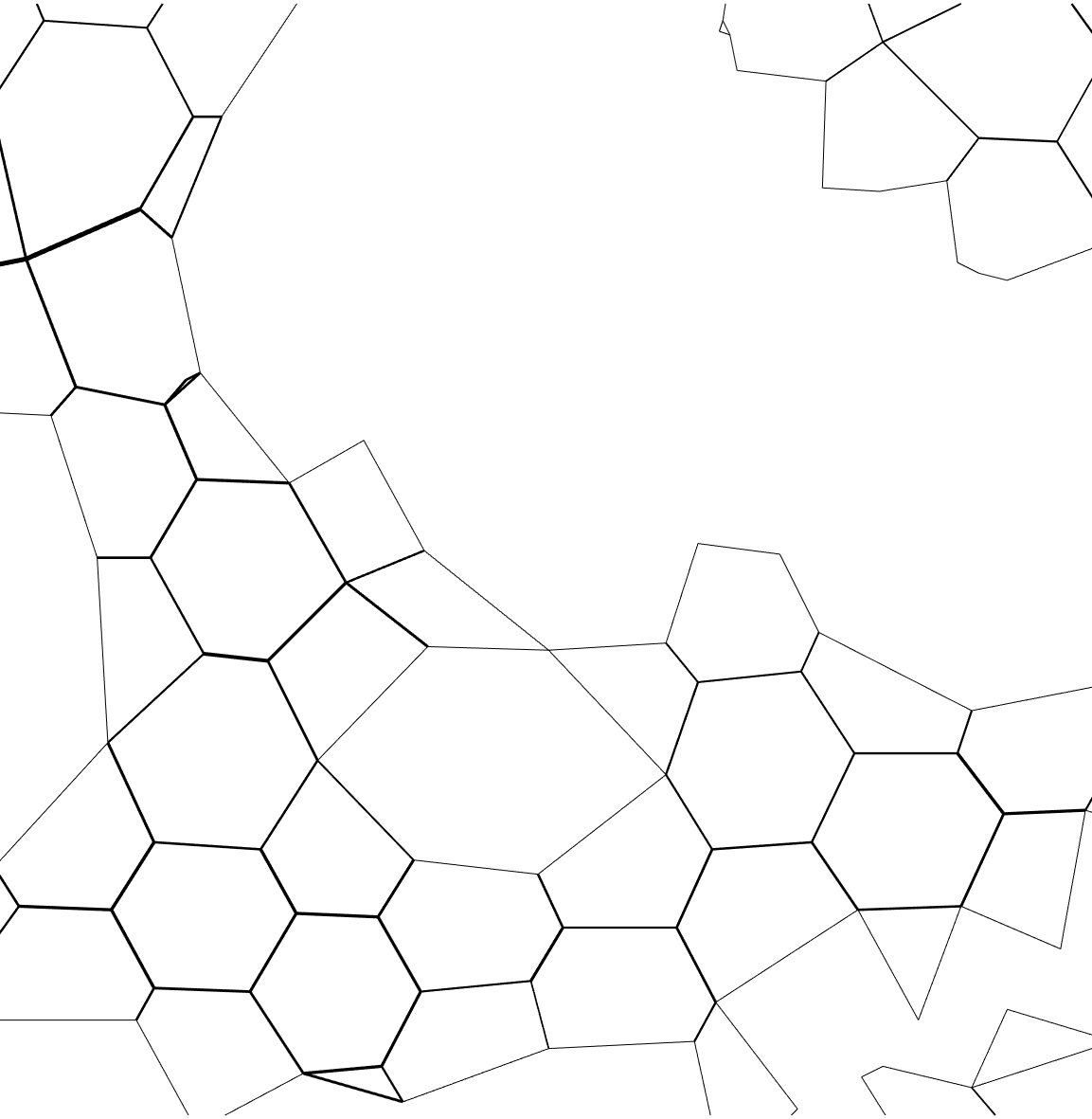} 
 \caption{Detailed network representation of the pore system. Taken in a Zoomed-in section of the $S_e=0.73$ case.
\label{fig:Detailed_net}}  
\end{wrapfigure}

The detailed representation of the pore system accounts for the
spatial location of the pores, their coordination number, and the
sizes of the pore bodies and throats
(Figure~\ref{fig:Detailed_net}). In order to generate the detailed
network, the segmented pore label image (attained from the image
analysis), in which each pore is assigned a different index (label),
is used to form a graph. The graph holds information about the
connectivity of each pore and its
neighbors (on an 8-connectivity directions basis). The site location is set equal to the location of the maximal
distance map value in the pore, i.e., to the center of the maximal
disk (Fig.~\ref{fig:seg}d). The bond length between each connected
pores couple is set as the distance between pore centers.
The pore geometry in the detailed network representation is parabolic,
see Subsection~\ref{sec:parabolic}.  

\section{Results and discussion}


In this section, we study the estimates of the different network models for permeability, hydraulic tortuosity, and the distributions of pressure and flow rate and compare them to the results of the direct number simulations, which represent the ground truth. These are the key quantities for the understanding of pore-scale flow and solute dispersion in saturated and partially saturated porous media~\cite{dentz2018,puyguiraud2021pore,ben2023solute}. 

\subsection{Permeability}
\begin{figure}
 \includegraphics[width=\textwidth]{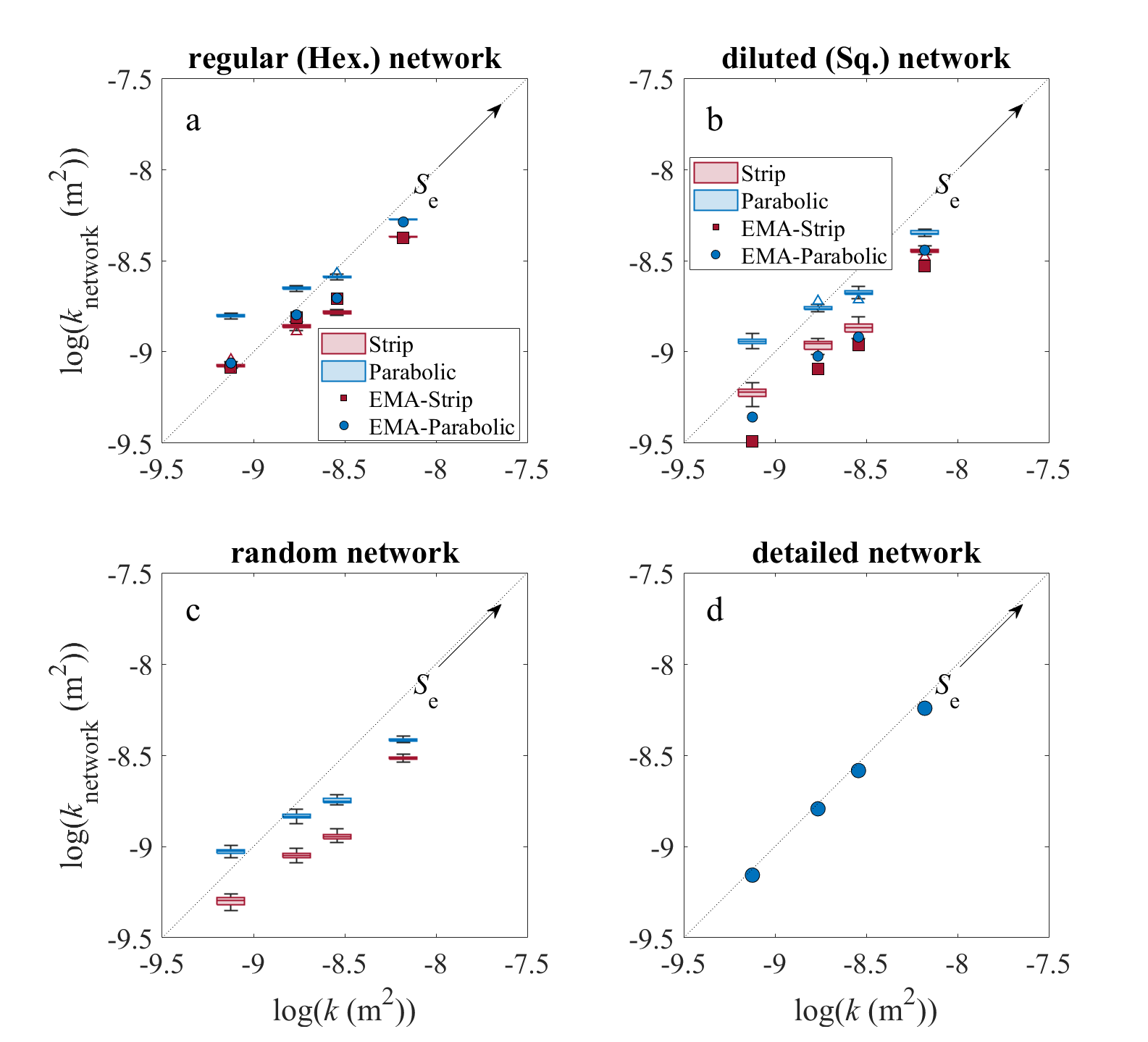}
 \caption{The permeability ($k$, Eq.~\ref{eq:permeability}) of the network compared to the numerically evaluated permeability for the a) regular, b) diluted, c) random, and d) detailed networks. 
 The different values of k are for the different saturation degrees under consideration (Fig. \protect{\ref{fig:media}}). 
 The boxplots of the regular (a), diluted (b), and random (c) models are evaluated from 25 realizations for each pore geometry (strip or parabolic). 
 The extreme values exceeding the $95\%$ confidence level (box) are depicted with triangles.}\label{fig:k}  
 \end{figure}

The ability to estimate the permeability ($k$) of a representative elementary volume (REV) is a basic prerequisite for using network models to upscale pore-scale flow phenomena to the Darcy and field scales. Figure~\ref{fig:k} shows the $k$ estimated from the different network models and $k$ based on the effective medium approximations (EMA) versus the permeabilities obtained from the direct numerical simulations (DNS). The estimates from the pore network models are based on averages of 25 network realizations for each pore geometry. The regular and diluted networks give a relatively narrow span of permeabilities for different realizations, suggesting that the domain is large enough to be considered a REV. As expected, for all network models, the permeability monotonically increases with the saturation degree $S_e$. 
Even if the permeabilities from the network models and direct simulation do not coincide, the network models may give a good estimate for the relative permeability $k_r = k/k_s$, where $k_s$ is the saturated (intrinsic) permeability. The effect of $S_e$ on $k_r$ for the different networks is presented in figure SF1 in the supplementary material.  

\paragraph*{Regular networks} \qquad
The regular networks (and also the diluted and random) estimate better the relative permeability ($k_r$, Fig. SF1) than the absolute one ($k$, Fig.~\ref{fig:k}). This is in accordance with the literature \cite{hunt2017}.
Here, the regular networks slightly underestimate the effect (slope) of $S_e$ on the relative permeability $k_r=k/k_s$ (Fig. SF1a).  
The increase in the standard deviations of the permeabilities (seen in the box plots in Fig.~\ref{fig:k}a,b) with the reduction of the saturation $S_e=1$ in the hexagonal grid is due to the changes in the pore size distribution with a broader distribution and many more small pores accounting for the narrow water films.

The $k$ evaluated by the EMA for the rectangular pore geometry (red rectangular markers in Fig.~\ref{fig:k}a) and parabolic (blue circle markers) pore geometries are very close. This is because of their relatively similar conductance probability density functions (Fig.~\ref{fig:gpdf}). As expected, the $k$ evaluated by the EMA is very close to the mean $k$ evaluated by the regular network models (box plots, Fig.~\ref{fig:k}a). Interestingly, the strip pore geometry (red) gives better agreement between the EMA and the network. Maybe because of the smaller variability in the bond conductances. 

\paragraph*{Diluted networks} \qquad

The diluted networks (Fig.~\ref{fig:k}b) provide a better estimate for $k_r$ as a function of $S_e$ than the regular networks (Fig. SF1b), but give worse estimates for $k$, mostly underestimating the data from the direct numerical simulations. The strip pore geometry in the diluted network (red boxes in Fig.~\ref{fig:k}b) presents higher variability between realizations (wider box plots) compared to the parabolic geometry. This is because of the combined effect of the larger fraction of low conductances (Fig.~\ref{fig:gpdf}a) and the network's dilution, which may form singular and dead-end sites. Interestingly, the $k$ values obtained from the diluted EMA are lower than the ones evaluated by the network models. 

\paragraph*{Random networks} \qquad

The estimates by the random networks (Fig.~\ref{fig:k}c)  for $k$ and k$_r$ are similar to those of the regular and diluted networks. This suggests that the added information held in the distribution of $Z$ as opposed to the mean coordination number does not improve the estimation of $k$.

\paragraph*{Detailed networks} \qquad

The permeability estimates of the detailed network model (Fig.~\ref{fig:k}d) are in excellent agreement with the direct numerical simulations. This confirms the ability of the detailed network model to represent the mean flow behavior through porous media. However, this model relies on detailed pore space information that may often not be available.

\subsection{Tortuosity}

\begin{figure}
 \includegraphics[width=\textwidth]{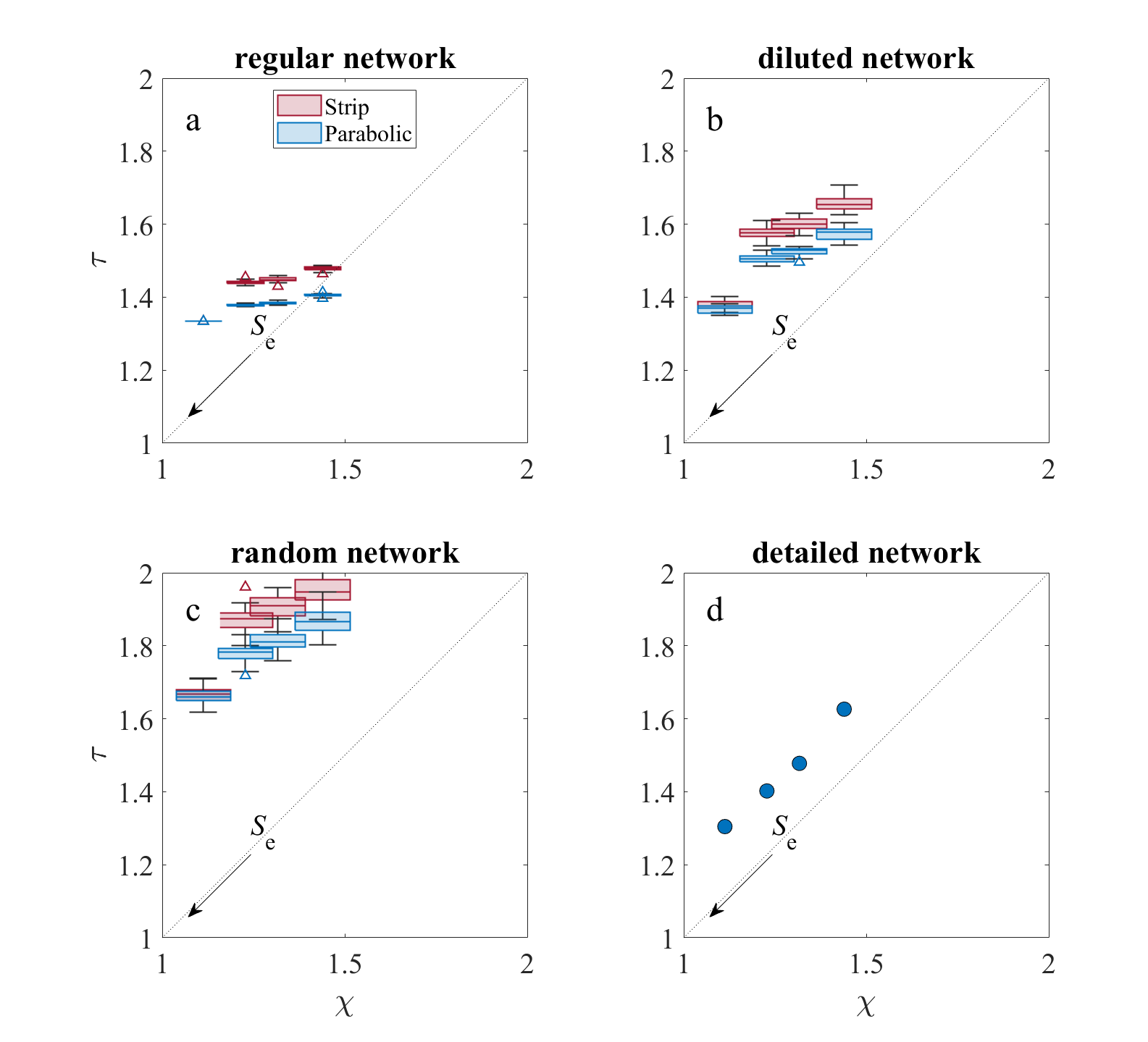}
 \caption{The hydraulic tortuosity on the network ($\tau$, Eq.~\ref{eq:tau}) compared to the numerically evaluated hydraulic tortuosity ($\chi$, Eq.~\ref{eq:chi}, for the a) regular, b) diluted, c) random, and d) detailed networks. 
 \label{fig:tau}}  
 \end{figure}

Figure~\ref{fig:tau} shows the hydraulic tortuosity estimates from the network models versus the tortuosity obtained from the direct numerical flow simulations using the digitized images of the porous media. Recall that the definition of tortuosity used for the network models in Eq.~\eqref{eq:tau} is different from the definition in Eq.~\eqref{eq:chi} for the direct numerical simulations. The two definitions coincide only if the bond lengths are constant, which is the case for regular networks. We observe for the networks and direct numerical simulations that tortuosity increases monotonically as the saturation degree decreases, which is due to the increasing complexity of the fluid phase configuration. Generally, the network models overestimate the values obtained from the direct flow simulations. A possible cause is the geometry of the lattice underlying the network models and its angle with respect to the pressure gradient. We define the relative tortuosity $\tau_r = {\tau}/{\tau_s}$, where $\tau_s$ is the tortuosity of the fully saturated case $S_e=1$. The effect of $S_e$ on the tortuosity can be evaluated from the slope of the markers or centers of the boxes compared to the dashed line in Figure~\ref{fig:tau}. It can be seen that all the models, except the regular networks, give a reasonable estimation of $\tau_r$. This suggests that the changes in the conductance distribution alone can not explain the increase of tortuosity with the reduction of $S_e$. 

\paragraph*{Regular networks} \qquad

In the regular lattice models,
tortuosity ($\tau$) originates from the geometry of the grid and from the distribution of the conductances. The conductances in this model are randomly assigned and not correlated. Thus, these models may underestimate preferential pathways and, therefore, slightly overestimate the actual hydraulic tortuosity $\chi$, see Figure ~\ref{fig:tau}a. 

\paragraph*{Diluted networks} \qquad

As outlined above, diluted regular networks may be characterized by singular regions, that is, disconnected subnetworks, and may also display stagnation points, such as dead-end pores. 
These features increase the hydraulic tortuosity so that the $\tau$ for the diluted networks (Figure ~\ref{fig:tau}b) are generally higher than for the regular networks (Figure ~\ref{fig:tau}a). 

\paragraph*{Random networks} \qquad
The random networks (Figure ~\ref{fig:tau}c) have the largest tortuosity, which suggests that the number of diluted bonds has a more significant effect on the tortuosity than the formation of stagnation and singular points. 

\paragraph*{Detailed networks} \qquad
The tortuosity estimated from the detailed network model gives the best approximation for $\chi$. This can be traced back to the effect of the dead-end regions on the preferential flow.

\subsection{Velocity and flow rate distribution}

The velocity and flow rate distributions play a key role in the understanding and prediction of solute transport from the pore and continuum scales. Variations in the micro-scale flow velocities give rise to the phenomenon of hydrodynamic dispersion \cite{bear1988}. The distribution of pore-scale velocities determines the evolution of solute dispersion and the tailing behavior of solute breakthrough curves~\cite{dentz2018,gouze2021modeling,ben2023solute}. The velocity distribution is a key property for the prediction of large-scale solute transport. 
It is not clear a priori that network models capture the velocity distribution observed in porous media because 1) networks do not account for the intra-pore velocity distribution, 2) networks do not reproduce the exact topological features of the pore space, and 3) the networks may display more stagnation sites, that is dead-end pores than one finds in the real medium.

\begin{figure}
 \includegraphics[width=\textwidth]{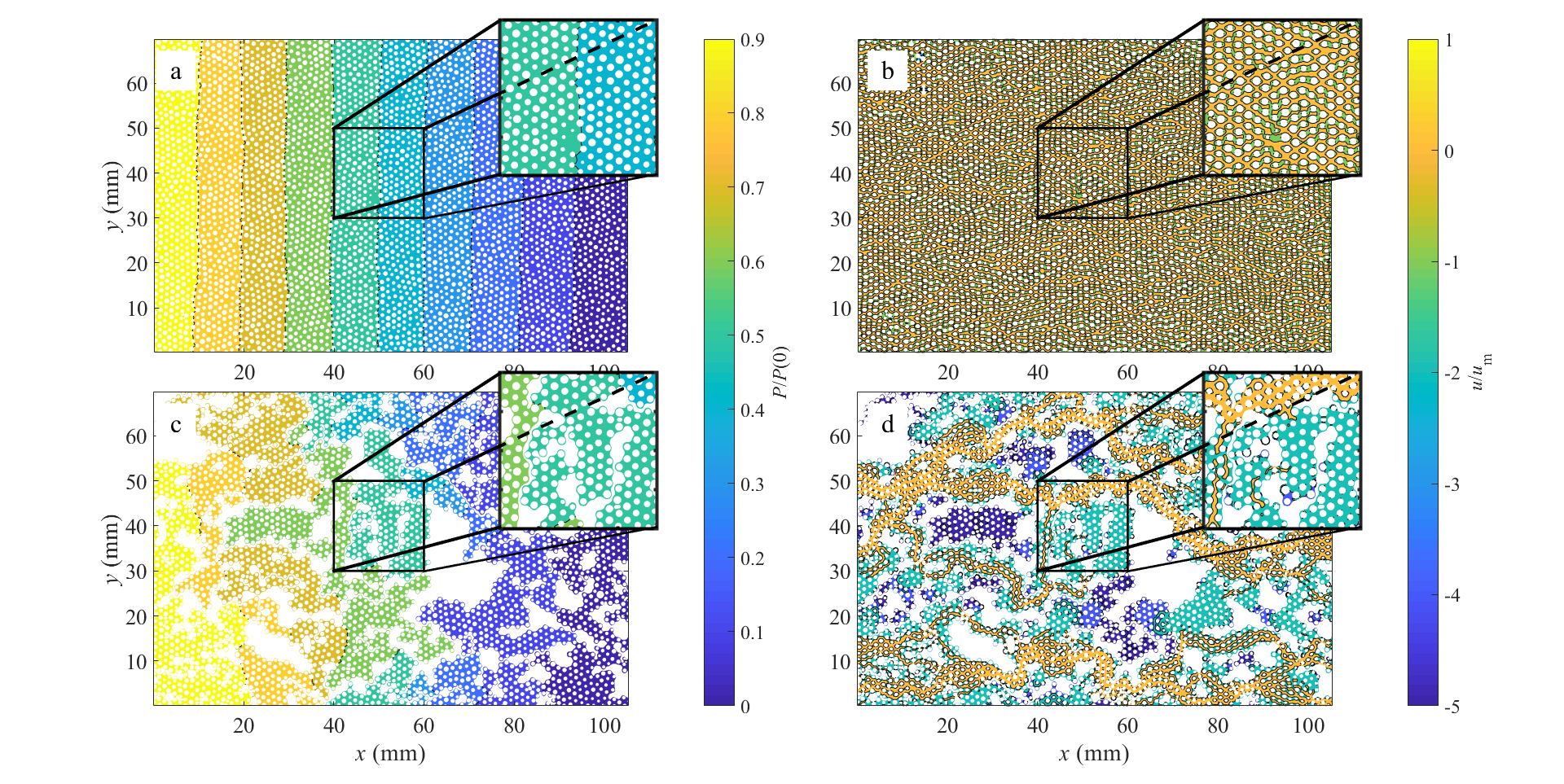}
 \caption{Spatial distribution of the water pressure (a and c) and flow velocity (b and d) for the fully saturated (a and b) and the $S_e=0.65$ (c and d) cases, evaluated by the numerical simulations. 
 \label{fig:numeric_P_and_v}}  
 \end{figure}

\begin{figure}
 \includegraphics[width=.9\textwidth]{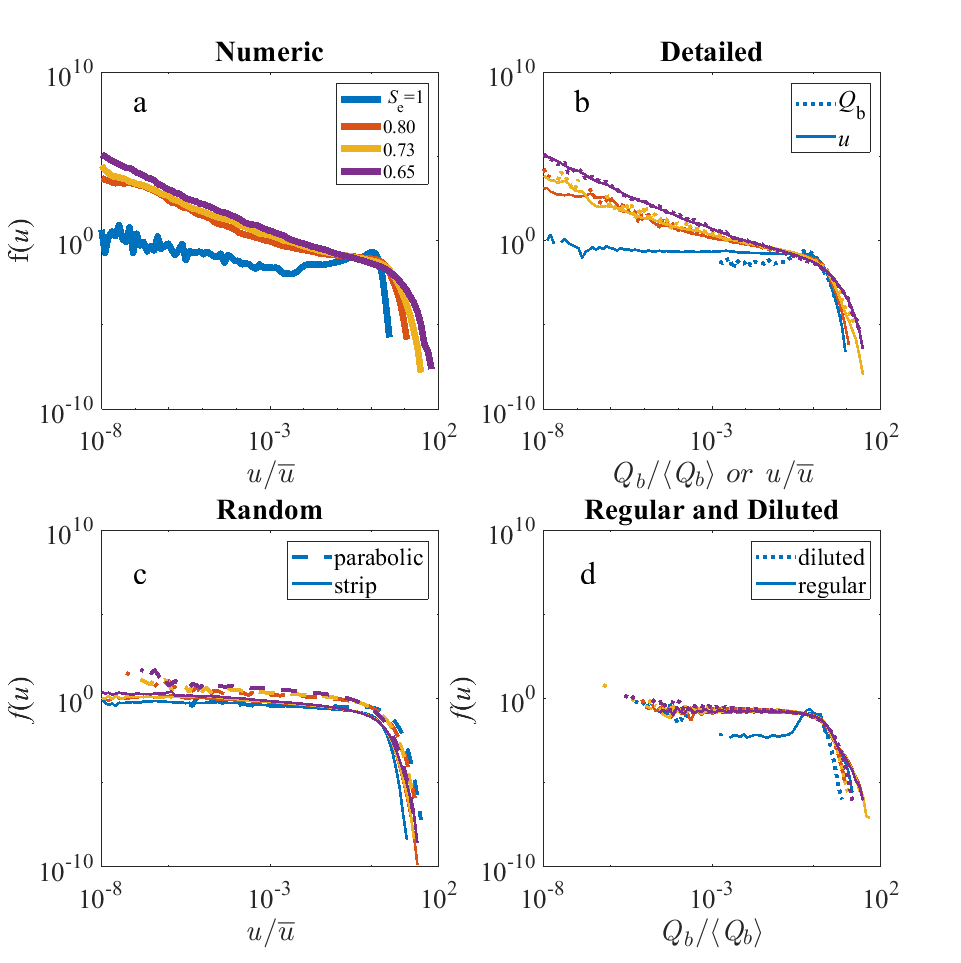}
 \caption{a) The velocity (normalized by the mean flow rate) probability density function (PDF) of the numerical simulation for the different saturation degrees ($S_e$), b) the flow rate at the bonds (Q$_b$) and the velocity distribution (u, Eq.~\ref{eq:u_distribution}) PDFs evaluated by the detailed representation of the pore space as a network model, c) the velocity PDF of the parabolic and strip pore geometries evaluated by the random network, and d) the bonds flow rate PDFs of the strip pore geometry evaluated by the regular and diluted network models. In these plots, $\Delta u_k$ intervals are chosen by dividing the logarithmic span of the velocities arbitrarily into 40 (equal on a logarithmic scale) intervals.}\label{fig:updf}
 \end{figure}

The effect of the saturation degree on the spatial pressure and flow distribution is depicted in Figure~\ref{fig:numeric_P_and_v} for the saturated case and the partially saturated medium with $S_e = 0.65$.  The saturated case has a relatively uniform flow velocity at the device scale and local low flow velocity regions between grains that are aligned in the main flow direction. For the partially saturated case in contrast, the local pressure gradients diverge from the main flow direction, causing a more tortuous flow. It is also visible that the unsaturated case has significant regions of low flow velocities and small pressure gradients. Figure~\ref{fig:updf}a shows the probability density functions of the flow velocity from the direct numerical flow simulations with flow velocities that range over almost ten orders of magnitude. The distributions are characterized by significant tails toward low velocities. As the complexity of the fluid-filled medium portion increases, the probability of low velocities increases, which is due to the low-velocity regions depicted in Figure~\ref{fig:numeric_P_and_v}. The low-velocity end of the velocity distribution determines the long-time tails of solute breakthrough and the evolution of dispersion. 

 \subsubsection{Regular, diluted and random networks} 
\begin{figure}
 \includegraphics[width=\textwidth]{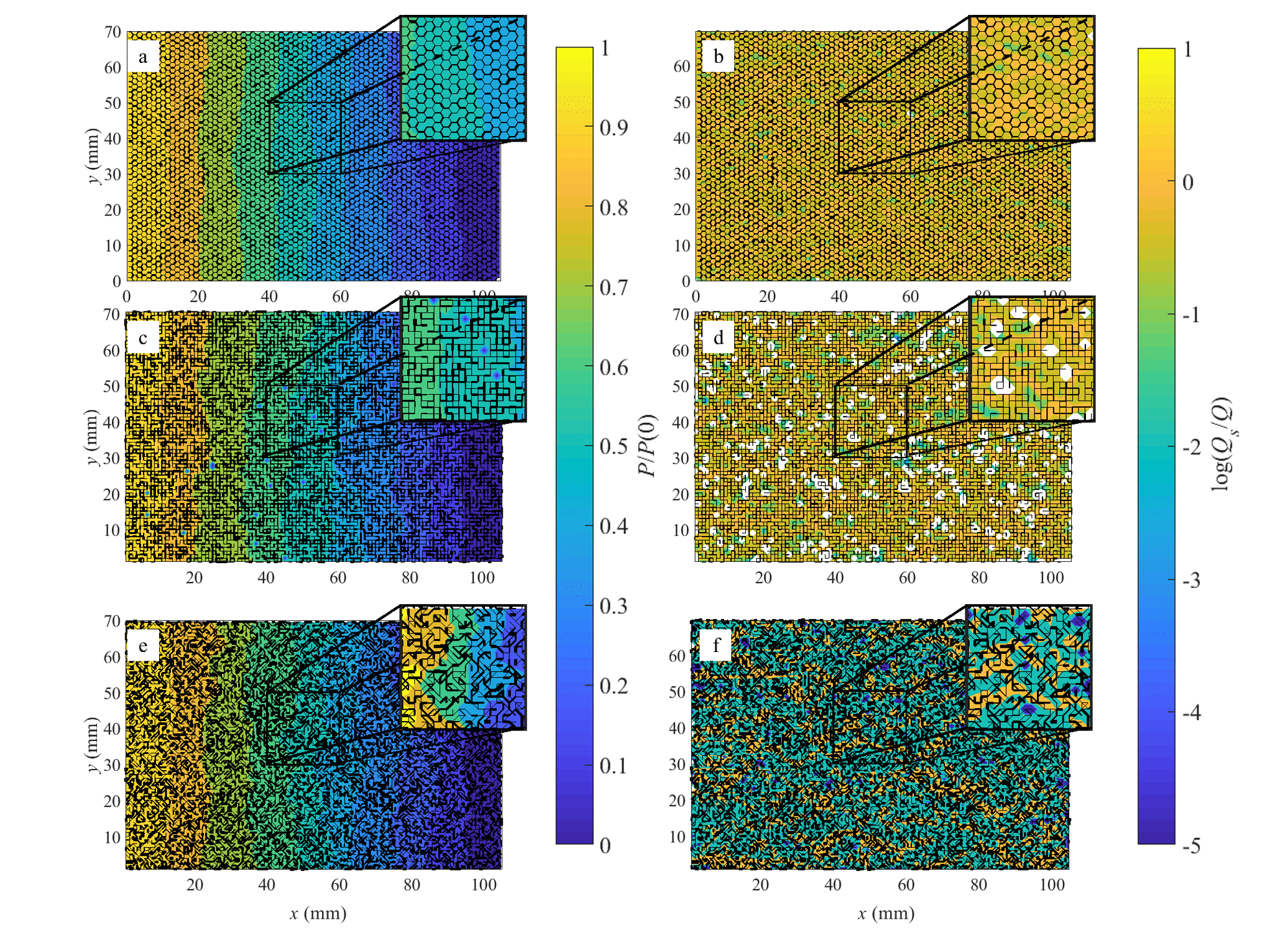}
 \caption{Spatial distribution of the pressure (P) at the sites (a, c, and e) and flow rate at the sites ($Q_s$, Eq.~\ref{eq:Q_s}) (b, d and f) for the regular (hexagonal) network (a and b), the diluted (square) network (c and d), and the random network (e and f) of the $S_e=0.65$ case with a parabolic pore geometry. The width of the black line represents the conductance (g) of each bond in the depicted realization.
 \label{fig:regular_diluted_P_and_Q}}  
 \end{figure}

Figure~\ref{fig:regular_diluted_P_and_Q} shows the spatial pressure and flow rate distributions for the medium with saturation $S_e =  0.65$ obtained from the (top panels) regular and (center panels) diluted and (bottom panels) random networks. The three network models have the same conductance distribution and mean coordination numbers of $\langle Z \rangle \approx 3$. The random network matches the distribution of coordination numbers of the porous medium.

Stagnation points, that is, sites of zero flow rate, are relatively frequent in the square lattice (diluted network, Figure~\ref{fig:regular_diluted_P_and_Q}d) but not very common in the hexagonal lattice (regular network, Figure~\ref{fig:regular_diluted_P_and_Q}b). This suggests that the orientation of the pore throats and not just the connectivity of the pore space is important for pressure and flow rate distributions. For the saturated case (not shown) both network models give spatial pressure and flow rate distributions that look qualitatively similar to those from the direct numerical simulations. Also, the velocity and flow rate PDFs estimated from the networks compare well to those obtained from the direct flow simulations (see Figures~\ref{fig:updf}c and d). For decreasing saturation degrees, however, the three network models are not able to describe either the pressure and flow rate maps or their PDFs. While none of the network models is able to reproduce the increase of low velocities with decreasing saturation, the random network picks up the wide range of flow velocities observed in the direct flow simulations. These observations indicate that the preferential flow patterns observed in the unsaturated medium in Figure~\ref{fig:numeric_P_and_v} cannot be explained by the variability in conductance or coordination number alone.   


For the diluted networks, the grid dilution to achieve the desired mean coordination number of $\langle Z \rangle = 3$ increases the number of stagnation and dead-end points compared to the regular square grid. These points give rise to the white regions in the $Q_s$ map (Figs.~\ref{fig:regular_diluted_P_and_Q}d), and blue regions in the pressure map (Fig.~\ref{fig:regular_diluted_P_and_Q}c). However, dilution also gives rise to a more pronounced preferential flow as compared to the regular network. The pressure distributions of the square diluted lattice (figure~\ref{fig:regular_diluted_P_and_Q}c) are more disordered than the regular hexagonal pressure distribution (figure~\ref{fig:regular_diluted_P_and_Q}a), for the same mean coordination number. This is because of the larger amount of dead-ends (white regions in figure~\ref{fig:regular_diluted_P_and_Q}b and \ref{fig:regular_diluted_P_and_Q}d) caused by the higher absolute dilution (starting from a higher saturated $Z$). This illustrates the limitation of a grid's mean coordination number to generally represent the medium's connectivity. 

The random network (Figures \ref{fig:regular_diluted_P_and_Q}e and f) shows a smaller number of singular points and dead-end pores than the diluted network (Fig.~\ref{fig:regular_diluted_P_and_Q}c and d), despite the larger degree of dilution (from $Z_{grid}=8$ for the base lattice to $\langle Z_{model} \rangle \approx 3$). This suggests that a wide distribution of the coordination number increases the pore-space connectivity. For illustration, consider the situation of ten pores, nine of which have a coordination number of $1$ and one of them a coordination number of $9$. Then, they are all connected.

In general, the size and spatial configuration of the stagnation areas are not well described by the lattice networks. This is to be expected because these network models (regular, diluted, and random) do not account for the spatial correlation of the porous space, only mean properties such as the mean coordination number and the one-point distributions of conductance and coordination number (only random network). In these network models, the changes in the saturation degree of the real media manifest in the total size assigned to the sites and bonds, or in other words, in the porosity represented by the network, that is, the ratio of the area of sites and bonds to the area of the medium. Note that the singular regions in the diluted networks are merely artifacts of the network generation process and do not contribute to the estimation of saturation of porosity. 
Moreover, the air phase is typically spatially correlated, as shown in Figure~\ref{fig:numeric_P_and_v}, which is not accounted for in these network models. Nevertheless, the random network models can reproduce some of the features obtained from the direct numerical simulations, such as the preferential flow paths (Fig.~\ref{fig:regular_diluted_P_and_Q}f) and the broad spectrum of flow in flow rates (Figure~\ref{fig:updf}). 

\subsubsection{Detailed networks} 
\begin{figure}
 \includegraphics[width=\textwidth]{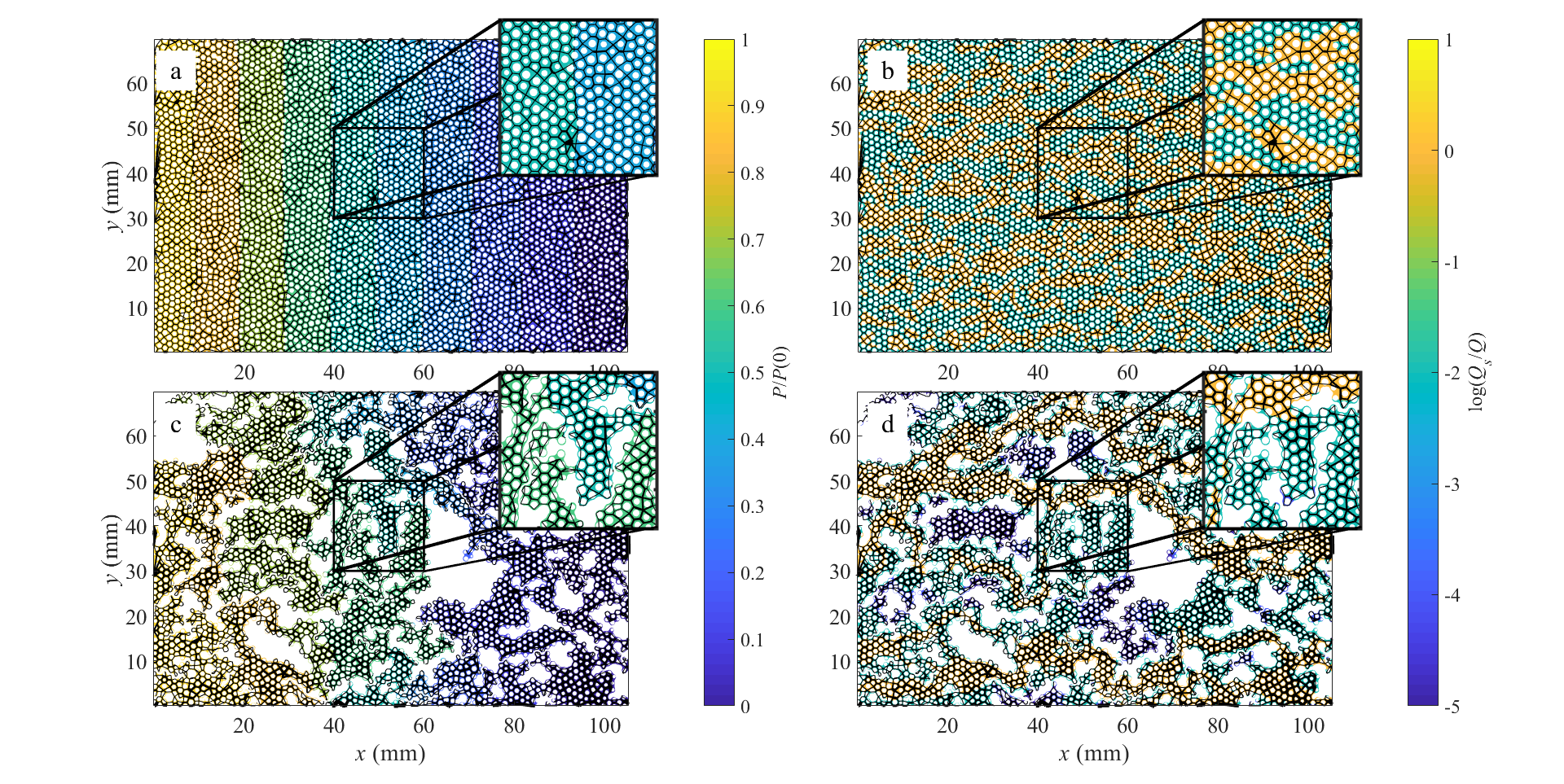}
 \caption{Spatial distribution of the pressure (P) at the sites (a and c) and flow rate at the sites ($Q_s$) (b and d) for the detailed network (c and d) for the $S_e=1$ (a and b) and $S_e=0.65$ cases with a parabolic pore geometry. The width of the black line represents the conductance (g) of each bond.
 \label{fig:rand_detailed_P_and_Q}}  
 \end{figure}

Figure \ref{fig:rand_detailed_P_and_Q} shows the pressure and flow rate maps for $S_e = 1$ and $0.65$ obtained from the detailed network representations of the media. The detailed networks show similar pressure and flow rate distributions as the numerical simulations (Figure~\ref{fig:numeric_P_and_v}) for the two saturation degrees. Also, the PDFs of flow rates closely match the ones obtained from the direct flow simulations, as shown in Figure~\ref{fig:updf}b. 
The detailed network model accounts only for the percolated connected fluid phase. Thus, it does not generate singular points or regions. In this model, the number of stagnation sites with zero velocity is small and comparable to the number of sites with $Z=1$. This is in accordance with the numerical simulations in which zero velocity prevails only at the solid-fluid interface and does not exist at any finite volume (mesh element). As seen in the map of flow rates both from the direct flow simulations (Fig.~\ref{fig:numeric_P_and_v}), the air phase encloses
extensive connected dead-end regions (of many pores) in which the flow is very slow. These regions do not participate in the conductance of fluids, which reduces permeability and increases preferential flow. Moreover, these regions give rise to non-Fickian solute transport due to diffusion-based solute entrapment and release \cite{liu2012, gjetvaj2015, ben2023solute}. All of these features are present in the detailed network models shown in Figure \ref{fig:rand_detailed_P_and_Q}. This observation shows the importance of accounting for the morphology of the connected fluid phase and pore space and demonstrates the viability of using network models to quantify not only the mean flow but also the flow distribution in porous media.

The detailed model is able to reproduce all features of the PDF of flow rates for the different media (Figure~\ref{fig:updf}b). It accounts correctly for the increase of low flow rates with decreasing saturation and captures the full range of flow rates. 
The flow rate PDFs are very similar in magnitude and shape to the volumetric flow rate PDFs, indicating that there is no large variability in the pore cross-sections. For the saturated case, however, the range of flow rates is much larger than the range of volumetric flow rates, indicating that the distribution of intrapore velocities is important.  However, with increasing heterogeneity of the pore space, i.e., with decreasing saturation, this difference vanishes and both flow rates and volumetric flow rates show a similar range and shape. Thus, the intrapore velocity distribution reproduction is less important for these cases.  

\section{Conclusion}
The conceptualization of a porous medium as a network of conductors is valuable and effective for quantifying flow and transport phenomena and evaluating the media's hydraulic properties. The amount of detailed information required for pore-space characterization depends on the complexity of the flow phenomenon and the flow and transport processes of interest. In this study, we demonstrated how different levels of information about the phase configuration in the pore space can be used to evaluate different properties of single-phase flow in variably saturated media.
The media permeability, even for the partially saturated cases, is well quantified even by relatively simple regular network models. For the saturated medium under consideration, the regular network model gives a good approximation of the hydraulic tortuosity and velocity distribution, albeit not for the velocity range. However, for the partially saturated cases, the more complex features, such as the media tortuosity and velocity distribution, are governed by the spatial configuration of the fluid phase. Thus, the accurate evaluation of these factors relies on a relatively detailed description of the pore space, its spatial distribution and connectivity. This information is available for the pore space in saturated porous media but not necessarily for the more complex fluid distribution and fluid-filled pore space in a partially saturated medium. Addressing this issue requires the quantification of the spatial configuration of the fluid phase and identifying its relevant quantifiers. 

\section*{Acknowledgments}
I.B.N, M.D. acknowledge funding from the European Union’s Horizon 2020 research and innovation programme under the Marie Sklodowska-Curie grant agreement No. HORIZON-MSCA-2021-PF-01 (USFT). I.B.N. , J.J.H. and M.D. acknowledge the support of the Spanish Research Agency (10.13039/501100011033), Spanish Ministry of Science and Innovation through grants CEX2018-000794-S and HydroPore PID2019-106887GB-C31. J.J.H. acknowledges the support of the Spanish Research Agency (10.13039/501100011033), the Spanish Ministry of Science and Innovation and the European Social Fund ``Investing in your future'' through the ``Ram\'on y Cajal'' fellowship (RYC-2017-22300). 

 \section{Open Research}
The data used in this work is provided as supplementary material for the review process and later will be made available in a data repository with a permanent identifier.

\appendix
\section{Effective medium approximation (EMA)} \label{sec:EMA}
The effective medium approximation (EMA) states that the network of variable
conductors can be represented by a regular uniform network with an effective
conductance $g_e$ . The EMA automatically obeys the
smooth field approximation \cite{Burganos1987}, i.e., the macroscopic
pressure field is a smooth function of position in the porous medium
\cite{Sahimi1990}, or in other words, the pressure gradient at each pore is
the projection of the macroscopic pressure gradient in the direction of the pore
axis \cite{Friedman1995}. The effective conductivity $g_e$ is 
calculated from the conductance distribution $f_g(g)$ by requiring that the
average of the flow perturbation caused by the EMA is zeroed
\cite{Kirkpatrick1973}
\begin{equation} \label{eq:EMA}
\int \frac{g-g_e}{g+(\frac{Z}{2}-1)g_e}f(g)dg=0 
\end{equation}
Then the EMA permeability of the media ($k_{EMA}$ [m$^2$]) can be evaluated from
$g_e$
\begin{equation} \label{eq:k_EMA}
k_{EMA}=\frac{\mu N_W}{bW}g_e
\end{equation}
where $N_W$ is the number of junctions on the lattice cross-section
perpendicular to the main flow direction. For a square grid
$N_W=\frac{W}{l_{sq}}$ while for the hexagonal case, where the hexagons are
oriented with the long diagonal in the direction of the flow,
$N_W=\frac{W}{\sqrt3l_{hex}}$. 

For the diluted network, we describe the conductance distribution in a binary form as a summation
of two terms \cite{Friedman1995}, a fraction $p$ of diluted bonds with
zero conductance and the fraction of the surviving bonds ($1-p$) so
that $f_1(g)=p\delta (g) + (1-p)f(g)$, where $\delta (g)$ is the Dirac
delta function and $f(g)$ is the conductance probability density
function (\ref{fig:gpdf}).

\bibliographystyle{elsarticle-num-names} 
\bibliography{cas-refs}

\end{document}


\begin{frontmatter}

\title{Pore network models to determine flow statistics and structural controls in variably saturated porous media}

\author[inst1, inst2]{Ilan Ben-Noah}
\author[inst1]{Juan J. Hidalgo}
\author[inst1]{Marco Dentz}

\affiliation[inst1]{Institute of Environmental Assessment and Water Research (IDAEA), Spanish National Research Council (CSIC), Barcelona, Spain.}
\affiliation[inst2]{Department of Environmental Physics and Irrigation, Institute of Soil, Water and Environmental Sciences, The Volcani Institute, Agricultural Research Organization, Rishon LeZion, Israel}

\end{frontmatter}

\noindent\textbf{Contents of this file}
\begin{enumerate}
\item Figure SF1
\end{enumerate}

%
%



%

 \begin{figure}[b] 
 \includegraphics[width=\textwidth]{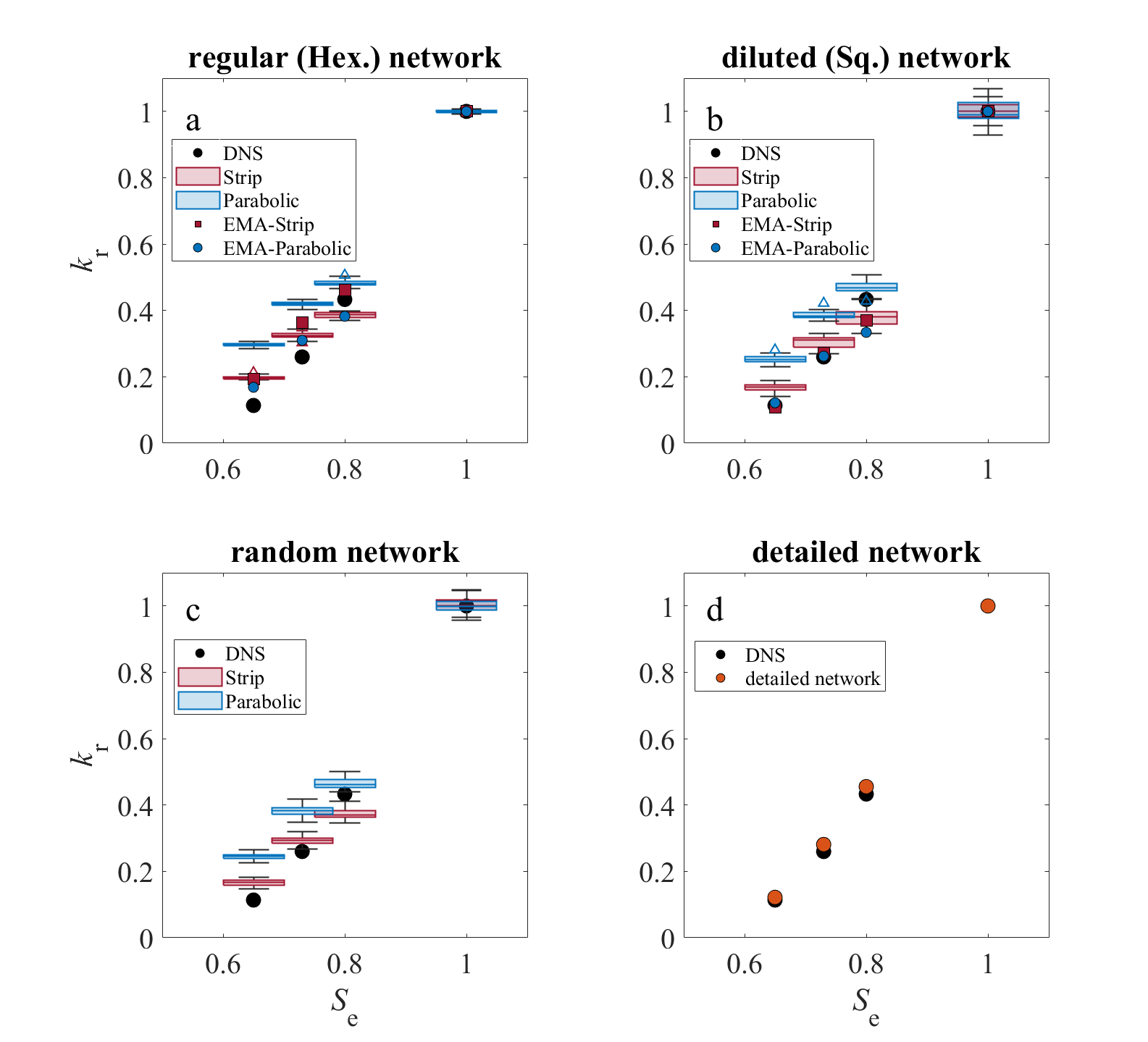}
 \caption{The effect of the effective saturation degree ($S_e$) on the relative permeability ($k_r$) for a) regular network, b) diluted network, c) random network, and d) detailed network for the different pore geometries (strip in red and parabolic in blue). The boxplots represent the statistics from 25 realizations, the direct numerical simulation (DNS) permeability values are in black circles.      
 \label{fig:k_r}}  
 \end{figure}